\documentclass[a4paper,10pt,oneside,showpacs,amsmath,amssymb]{article}
\usepackage[cp1251]{inputenc}
\usepackage{mathtext}
\usepackage{cite}
\usepackage{amssymb}
\usepackage{indentfirst}
\usepackage{caption}
\usepackage{graphicx}
\usepackage[left=20mm,top=20mm,bottom=20mm,right=15mm]{geometry}
\usepackage{amsmath}

\begin{document}
\title{Rate theory modeling a vacancy mediated intra-granular  fission gas bubbles growth in amorphous $U_3Si_2$}

\author{V.O.Kharchenko$^{a}$, D.O.Kharchenko$^{a,b}$\footnote{Corresponding author: dikh@ipfcentr.sumy.ua}, O.M.Shchokotova$^a$, V.V.Kupriienko$^a$, Lu Wu$^b$\\
$^{a)}$~Institute of Applied Physics, National Academy of Sciences of Ukraine\\
$^{b)}$~ Nuclear Power Institute of China
}
 
%\date{}
%
\maketitle
\begin{abstract}
A model  for gas bubble behavior in irradiated amorphous $U_3Si_2$ is generalized to take into account local influence of sinks for point defects and gas atoms as far as defect clustering resulting in growth of dislocation loops. A universality of bubble size distribution function and scaling law of bubble size growth is revealed. Temperature dependencies of main quantities governing bubble growth are discussed. Local distribution of bubbles and dislocation loops inside grains is studied in details to illustrate bubble size change in the vicinity of grain boundaries and estimate local swelling. Obtained data are compared with experimental and numerical studies.   
\end{abstract}

%\graphicspath{{pdf_figs/}}
\section{Introduction}

Challenges to  enhance of both the safety and performance of Light 
Water  Reactors (LWRs)  with $UO_2$ as nuclear fuel remain  an actual research  area in last several years.
It is known that $U_3Si_2$ can be  alternative to $UO_2$ as accident tolerant fuel  in commercial LWR (low temperature application). By comparing it with  uranium dioxide one can issue   higher
uranium density for $U_3Si_2$  (17\% more uranium than $UO_2$), increasing power efficiency, and  better
thermal conductivity (8.4 $W/m.K$ at 338 $K$ \emph{versus} 6 $W/m.K$ at 400 $K$), resulting in a safety enhancement. Disadvantages of this silicide are in  a lower melting
temperature (1938 $K$ \emph{versus} 2847 $K$) and increased water reactivity \cite{USi2020}.

Studies related to this material behavior at typical reactor conditions are provided experimentally in research  and test reactors (for, example in Oak Ridge (ORR), Idaho National Laboratory (ATR), SCK CEA (BR2, OSIRIS).
Fuel pellets of $U_3Si_2$ are usually fabricated by 
arc melting, atomization, and subsequent sintering \cite{HOP94}.
It has been
used as a dispersed phase in aluminium plate-type reactor fuel. The most of irradiation data illustrate  radiation behavior under relatively low temperatures
(<500$K$) with high fission density \cite{Exp2009,Exp2009_1,Exp2010}.

Despite several advantages of $U_3Si_2$ comparing to $UO_2$ aspects related to its swelling behavior remain under discussion. It is related to a differences in swelling mechanism in research reactors and commercial LWR caused by amorphization scenario: fission-induced amorphization affecting the swelling \cite{Finlay04} can be suppressed at much high operating temperatures in LWRs (see Refs.\cite{MKW2014,NBAMC2016,MiaoJNM17_484,MiaoJNM17_495,MiaoJNM18_503,
YaoJNM18_498,MiaoJNM18_511,AABM2020}).

It is well known that fission gas behavior acts crucially onto 
fuel performance. Gaseous fission products, primarily Xe,
aggregate within the fuels and form intra- and intergranular
bubbles \cite{Report2017}. It results in an  increase the
porosity of the fuel materials, leading to gas swelling
and a reduction in thermal conductivity \cite{Tonks2016}.
It was  shown that under research reactor conditions small ($<1nm$) bubbles can not be detected by scanning electron microscopy until a threshold for fission density is achieved. Visible bubbles emerge above the threshold when one can observe  breakaway swelling known as ``knee'' phenomenon \cite{Exp2009,Exp2009_1}. It relates to the regime where the interconnection of nanometer size bubbles occurs in bubbles grow as a result of the coalescence. 

Experimentally it was shown that  all these bubbles are of spherical form with density around $10^{13}-10^{17} cm^{-3}$ depending on studied conditions  \cite{Rest2004,Finlay04,MKGAYMHY2017}. It was shown in ORR testing at temperatures below 384 $K$ (athermal bubble growth) that  $U_3Si_2$ is more stable to swelling comparing to other silicides \cite{USi2020}. $U_3Si_2/Al$
plate-type dispersed fuel manifests uniformly-distributed separate small-sized fission gas bubbles after a burn up of $5.2\times10^{21} fission/cm^3$ \cite{Finlay04}. A the same time a temperature governed bubble morphology was found in $U_3Si_2/Al$ plate fuel
irradiated at Advanced Test Reactor \cite{Exp2009}.

Among experimental and computation approaches devoted to qualify $U_3Si_2$ as accident tolerant fuel for LWRs the developed rate theory models were used to develop   GRASS-SST code for modeling fission gas behavior \cite{GRASS-SST}.
By using GRASS-SST simulation it was shown that at temperature ranges from 390$K$ (research reactor temperature) up to 1000$K$  the gas swelling is controlled by intragranular bubbles\cite{MKGAYMHY2017}. Analysis of bubble size distribution has shown that small bubbles are of the size around 1.5nm at low temperatures ($<750K$). Large  intragranular bubbles with the size up to 20nm emerge at elevated temperatures due to 	enhanced fission gas (Xe) diffusion leading to bimodal bubble size distribution. At burn-up around 6\% this distribution is unimodal. This conclusion came from results of theoretical studies \cite{Rest2004} and experimental observations at Xe-ion irradiation with the growing mean bubble size from 2nm up to 5nm  at doses from 20dpa up to 80dpa (see Ref.\cite{YaoJNM18_498} and citations therein). 
GRASS-SST simulations were used especially to study gas behavior in LWRs and in comparison to conventional oxide LWR fuels \cite{WD2017}.
It was pointed out that these studies are  heavily
relied on the experimental data from reactors \cite{MKGAZA2018}. Behavior of fission gas under  various loss-of-coolant accident  (LOCA) conditions  (at temperatures from 600K up to  1600K) in 
LWRs was simulated using rate theory (GRASS-SST code) with thermomechanics  (BISON code \cite{BISON}) , where it was shown that if the temperature is limited beneath 1200 K throughout the LOCA, the corresponding
fission gas behavior is expected to be benign   \cite{MKGAZA2018}.
 Above 1000$K$ up to accident scenario temperature (1190$K$) the grain boundary bubbles are formed. Here they start to play a major role in the swelling. With a further temperature increase intragranular bubbles interconnect and gas release channels 	are formed, resulting to dramatic drop of the swelling.  
According to obtained simulation results one issues that under normal LWR conditions  the overwhelming majority of fission gas is retained
in intragranular bubbles.

In this paper we follow previous studies of  bubble size dynamics in $U_3Si_2$ fuel elements  by using reaction rate theory and discuss thermal effects and local distribution of bubbles in grains by considering moderate temperatures. We generalize  models proposed in Refs.\cite{Rest2000,Rest2004} by taking into account local distribution of point defect and gas atoms sinks, namely, a local change of sinks (grain boundary) preference depending on the  distance from grain boundary, as was shown in rate theories (see Refs.\cite{RRT,Singh97}) and in phase field theory (see, for example Ref.\cite{LHSS2017} and citations therein).  It will be shown that both  interstitial loops as bubbles are distributed locally inside grains. It will be shown that bubble size distributions are characterized by  heavy tails mostly in  grain center domains, whilst these distributions are cut in areas located nearby grain boundaries where mostly small bubbles emerge. A developed formalism to study bubble size distribution is based on formalism discussed in Refs.\cite{Speight71,Rest2004}. In our study we consider so-called auto-model regime allowing to get time dependent bubble size distribution functions and obtain directly scaling dynamics of mean bubble size and their number density at the ``knee'' when interaction of bubbles results in a coalescence regime.

The paper is organized in the following manner. In Section 2 we briefly describe a model and introduce modification of rate theory model introducing defect clustering and dislocation loop size depending preference factors for point defects and gas atoms with local distribution of sinks inside grains. At the end of the section we proposed a receipt to obtain stationary bubble size distribution and time dependent distribution where one takes into account that bubbles are composed by gas atoms and vacancies. 
Section 3 is devoted to discussing main results of rate theory modeling. Here we analyze protocols of obtained data at different conditions and consider temperature dependencies of mean bubble size and mean dislocation loop size. At the end of this Section we  analyze  bubble size distribution functions. We conclude in Section 4.
     
\section{Model}
\subsection{General rate equations}

Following the approach developed in Ref.\cite{Rest2004} we describe dynamics of interstitial and vacancy  concentrations ($c_i$, $c_v$) by following equations:
\begin{equation}\label{eqpd}
\begin{split}
\partial_t c_i&=K_i-k^2_iD_ic_i-\alpha_r c_ic_v-\frac{\sqrt{2}}{\Omega_0^{2/3}}D_ic_i^2+\frac{\Omega_0^{1/3}b_vD_v}{\pi }\frac{N_Ic_v}{R_I},\\
\partial_t c_v&=K_v-k^2_vD_vc_v-\alpha_r c_ic_v-\frac{\Omega_0^{1/3}b_vD_v}{\pi }\frac{N_Ic_v}{R_I}.
\end{split}
\end{equation}
Defect generation rate $K_q=\Omega_0 G\dot f(1-\varepsilon_r)(1-\varepsilon_{qg}-\varepsilon_{qs})$ is defined through fission rate  $\dot f$, number of survived Frenkel pairs per fission $G$, atomic volume $\Omega_0$, efficiency of cascade relaxation at fission  $\varepsilon_r$, fraction of sessile clusters $\varepsilon_{qs}$ and glissile clusters $\varepsilon_{qg}$, where $q=\{i,v\}$. $D_i$ and $D_v$ are point defect diffusivities, $b_v$ is the Burgers vector modulus, $\alpha_r$ is the recombination rate constant defined in the standard manner, $N_I$ and $R_I$ are interstitial dislocation loop number density and radius, respectively.

Sink strengths of point defects are: $
k_q^2=Z_{Nq}\rho_N+Z_{Iq}\rho_I+Z_{bq}R_bN_b +Z_{GB}\lambda_{GB}^{-2}$,
where $Z_{Nq}$ defines preference factors for network dislocations with  density $\rho_N$, $Z_{Iq}$ relates to dislocation loops (of interstitial character) of the density $\rho_I=2\pi R_IN_I$, $Z_{GB}$ corresponds to  grain boundaries of the radius $\lambda_{GB}$, $Z_{bq}$ describes effect of  bubbles having the size $R_b$ and  number density $N_b$. In our case we put
$
Z_{Ni}=1.1$, $Z_{Nv}=1.0$; $Z_{Iq}={2\pi}\left[\ln({8R_I}/{r_{0q}})\right]^{-1}$, where $r_{0q}$ denotes loop-defect interaction radius;  $Z_{Ii}>Z_{Iv}$,
$Z_{qb}=4\pi Y_{q}$, where $Y_q$ relates to preference of absorption of $q$-type point defects \cite{HSB2020}, in our case we put $Y_q=1$ for simplicity;
$Z_{GB}=\frac{3\theta^2(\theta\coth(\theta)-1)}{\theta^2-3(\theta\coth(\theta)-1)}$, $\theta=\sqrt{\rho_N}\lambda_{GB}$.  
To describe an influence of grain boundary onto bubble-size distribution inside grains we take into account local distribution of dislocations inside grains as was shown in Ref.\cite{DSW03}. The approximation formula describing this locality is as follows: $\rho_N=\rho_{N0}e^{-\ell_0/\ell}/(1+e^{-\ell/\ell_0})$, where $\ell=l/\lambda_{GB}$, $l$ is the distance from grain boundary measured in units $\lambda_{GB}$ where $\ell_0=0.001$ is a fitting parameter describing a length free of dislocation cores nearby grain boundaries \cite{DSW03}.  

Equations for loop number density and radius take the form  \cite{Rest2000}
\begin{equation}
\partial_t N_I=\frac{\sqrt{2}}{2}\frac{D_i}{\Omega_0^{5/3}}c_i^2-\frac{4N_I}{\pi R_I b_v}\left(Z_{Ii}D_ic_i-Z_{Iv}D_vc_v\right)-\frac{b_vD_v}{\pi\Omega_0^{2/3}}\frac{c_vN_I}{R_I},
\end{equation} 
\begin{equation}
\partial_tR_I=g_I\left(b_v^{-1}\left[Z_{Ii}D_ic_i-Z_{Iv}D_vc_v\right]-(R_I-R^0_I)\frac{\partial_tN_I}{2N_I}+\frac{K_i\varepsilon_{ig}(\pi r_{icl})^2\tilde\rho}{2d_ib_vS_{icl}(\ell)}\right)
\end{equation}
Here $R_I^0$ is a cut off for loop radius, $\varepsilon_{ig}$ stands for a fraction of glissile interstitial clusters.  We take into account that  the sink strength of defect clusters, $S_{icl}$,    depends on the distance \cite{Singh97,RRT}:
$S_{icl}(\ell)=2\left( {\pi r_{icl}\tilde \rho}/{2}\right)^2,$ $
\tilde\rho=\rho_N+\rho_I+{2}[{\pi r_{icl}}{\sqrt{l(2\lambda_{GB}-l)}}]^{-1}$, 
where  $r_{icl}$ is the capture radius of  loops for mobile clusters, interstitial clusters can migrate in $d_i=3$ crystallographic directions assumed equal.
The cellular dislocation structure is taken into account by introducing loops growth limitation factor \cite{Rest2000}: 
$
g_I=C_AC_\rho\sqrt{{\pi}/{f(\nu)}}-2R_I\sqrt{\tilde\rho}$, $C_A=3$, $C_\rho=2$, $f(\nu)=({1-\nu/2})/({1-\nu})$, where $\nu$ is the Poisson ratio.

It is assumed that network dislocation density can be produced and relaxed. To this end we exploit a  model  proposed in Ref.\cite{Stoller90}: $\partial_t\rho_N=2\pi(v_{cl}S_{BH}+r_{max}N_I/\tau_{max})-\rho_N/\tau_{cl}$. Here $v_{cl}=b_v^{-1}(Z_{Ii}D_i\overline{c}_i-Z_{Iv}D_v\overline{c}_v)$ is the climbing velocity, $S_{BH}=(\rho_p/3)^{3/2}$ is the Bardeen-Herring source density, $\rho_p$ is the pinned dislocation density, $\tau_{cl}=d_{cl}/v_{cl}$ is the climbing time, $d_{cl}=(\pi\rho_N)^{-1/2}$ is the climbing distance, $r_{max}=(\pi(\rho_N+\rho_I))^{-1/2}$ is the limitation for the maximal loop size, $\tau_{max}=r_{max}/v_{cl}$ is the corresponding maximal time until loop of maximal size will annihilate with dislocations.
By taking into account that pinned dislocations are fraction $\varpi$ of network dislocations and treated as adjustable parameter we arrive at the equation of the form 
\begin{equation}
\partial_t\rho_N=b_v^{-1}(Z_{Ii}D_i\overline{c}_i-Z_{Iv}D_v\overline{c}_v)\left(2\pi \left[(\varpi\rho_N/3)^{3/2}+N_I\right]-\sqrt{\pi}\rho_N^{3/2}\right).
\end{equation}
In our modeling we put $\varpi=0.1$ \cite{Stoller90}.

Fission gas atom (Xe) concentration inside grains obeys the equation \cite{Rest2004}  
\begin{equation}
\partial_t c_g= K_g-16\pi \alpha_0\eta_0 r_gD_gc_g^2-4\pi D_gR_bN_bc_g+ 2 b m_bN_b-Z_{GB}D_gc_g\lambda_{GB}^{-2}.
\end{equation}
The generation rate of gas atoms is $K_g=\beta \dot f$, $\beta$ is gas atom production per fission; $b$ is gas-atom resolution rate, $\eta_0$ is the viscosity, $r_g$ is  gas atom radius, $\alpha_0$ is the proportionality constant.
Gas atom radiation enhanced diffusion coefficient we exploit formula \cite{Rest2000}
$
D_g=D_g^0+D_g^{RED}\dot f$, $D_g^0=[{6\pi r_g\eta_0}]^{-1}$,  
$D_g^{RED}$ is radiation enhanced diffusion used to relate data to experiments.

The bubble number density obeys the equation \cite{Rest2004}
\begin{equation}
\partial_t N_b=16\pi\alpha_0\eta_0 r_g\frac{D_g c_g^2}{m_b}-bN_b-16\pi R_bD_bN_b^2, 
\end{equation}
where the gas-bubble diffusivity is $
D_b={3\Omega_0 D_g}/{4\pi R_b^3}$.

 As far as gas bubbles contain both gas atoms and vacancies,  the bubble mean size, $R_b$,  can be found from the bubble volume $V_b=\Omega_g m_b+\Omega_v n_{vb}$ as  $R_b=(3V_b/4\pi)^{1/3}$, where  $\Omega_g$  and  $\Omega_v$ are  gas atom volume and vacancy volume, respectively,  $m_b$ and $n_{vb}$ are mean numbers of gas atoms and vacancies in gas bubbles. 

Equation for gas atoms and vacancies  in bubbles are of the form  \cite{Rest2004}:
\begin{equation}\label{eqmn}
\begin{split}
\partial_t m_b&=4\pi R_bD_gc_g-bm_b+16\pi R_bD_bm_bN_b,\\
\partial_t n_{vb}&=(Z_{bv}D_vc_v-Z_{bi}D_ic_i)\frac{R_b}{\Omega_v}+\frac{2\pi D_{v} d}{k_BT}(p-p_{eq}).
\end{split}
\end{equation}
The gas-atom resolution rate is proportional to
the fission rate as $b=b_0\dot f$.
One assumes that intra-granular bubbles are spheres. The equilibrium pressure of the gas babble is $p_{eq}=2\gamma/R_b$, $\gamma$ is the surface tension. During accumulation of gas atoms and vacancies from matrix  the gas pressure in bubble is defined as 
\cite{Barani2019}: $p={T m_b}/{\Omega_v n_{vb}}$, here it is assumed that gas atoms are not in clusters.
The corresponding volume change due to gas local swelling is defined in the standard manner as $
{\Delta V (\ell)}/{V}=N_b(\ell) V_b(\ell)$. The total swelling is obtained as a sum over all possible configurations inside grain, i.e. $({\Delta V}/{V})_{tot}=\sum_\ell N_b(\ell) V_b(\ell)$. A fission density we measure as $F_D=\dot f t$. By solving the system of dynamical equations (\ref{eqpd})-(\ref{eqmn}) one gets time (fission density) dependencies used in computation of parametric dependencies of measured quantities and calculation of bubble size distribution functions. 

\subsection{Distribution function over bubble size}
To find a mathematical construction for bubble size distribution function  $\Phi(R_b)$ we  follow the  receipt described in Refs.\cite{Speight71,Rest2004} and neglect diffusion of bubbles by considering only major mechanisms in their evolution. 
Let $\Phi(R_b){\rm d}R_b$ be the number of bubbles per unit volume with radii between $R_b$ and $R_b+{\rm d}R_b$.  
A growth and resolution of bubbles due to irradiation  is described by:   
\begin{equation}
\left[\frac{{\rm d}\Phi}{{\rm d}t}\right]_{g}{{\rm d}R_b}=-\frac{{\rm d}}{{\rm d}R_b}\left[\Phi(R_b)\frac{{\rm d}R_b}{{\rm d}t}\right]{\rm d}R_b,\quad \left[\frac{{\rm d}\Phi}{{\rm d}t}\right]_{r}=-b_0\dot f\Phi(R_b). 
\end{equation}

The total net of  rate of change of concentration of bubbles is 
\begin{equation}\label{eqPhi}
\left[\frac{{\rm d}\Phi}{{\rm d}t}\right]_{tot}=\left[\frac{{\rm d}\Phi}{{\rm d}t}\right]_{g}+\left[\frac{{\rm d}\Phi}{{\rm d}t}\right]_{r}.
\end{equation}
An equilibrium bubble population is defined in stationary case resulting in
\begin{equation}
\frac{{\rm d}\Phi(R_b)}{{\rm d}R_b}v(R_b)+\Phi(R_b)\frac{{\rm d}v(R_b)}{{\rm d}R_b}+b_0\dot f\Phi(R_b)=0,
\end{equation}
where the bubble growth speed  
\begin{equation}
v(R_b)=\frac{1}{4\pi R_b^2}\left(\Omega_g\frac{{\rm d}m_{b}}{{\rm d}t}+\Omega_v \frac{{\rm d}n_{vb}}{{\rm d}t}\right)
\end{equation}
is defined from the bubble volume $V_b=4\pi R_b^3/3$. 

A solution of equation for distribution function can be found in quadratures:
\begin{equation}
\Phi(R_b)=\frac{C_0}{v(R_b)}\exp\left(-b_0\dot f\int_{R_b}\frac{{\rm d}R}{v(R)}\right),
\end{equation}
where $C_0$ is the integration constant giving the total number of bubbles at the knee (the knee in the swelling curve it is ascribed to the regime where interconnection between nanometer size bubbles occurs) 
\begin{equation}\label{Nb}
C_0^{-1}\equiv N_{b}^{knee}=\int\limits_{R(0)}^\infty\Phi(R_b){\rm d}R_b.
\end{equation}
The average size of bubbles by this distribution is obtained in the standard way: $
\langle R_b\rangle_\Phi=%\frac{1}{N_b^{knee}}
\int_{R(0)}^\infty R_b\Phi(R_b){\rm d}R_b$.

A nonstationary solution of the equation for  the distribution function  $\Phi(R_b,t)$ can be found in auto-model regime. To this end we exploit the Fourier method and  put  
\begin{equation}
\Phi(R_b,t)=\phi(t)\varrho\left(y\right), \quad y=\frac{R_b(t)}{a(t)},
\end{equation} 
where $y$ is the bubble radius scaled in $a(t)$, $\phi(t)$ and $\rho(y)$ are unknown functions. The time derivative of the bubble radius defines the bubble growth speed ${\rm d}R_b/{\rm d}t=\theta(t)\omega(y)$, where the corresponding time dependence defined by $\theta(t)$ should coincide with $\dot a(t)$ following the definition of the bubble size derivative, i.e., $\theta(t)\equiv \dot a(t)$.

By taking  time and spatial derivatives  in Eq.(\ref{eqPhi})
%and spatial derivative as ${\rm d}\Phi/{\rm d}t=\varrho(y)\dot \phi$, ${\rm d}R_b/{\rm d}t=\omega(y)\dot a$ and ${\rm d}/{\rm d}R_b={\rm d}/a{\rm d}y$ 
%instead of Eq.(\ref{eqPhi}) 
one arrives 
%can  separate parts describing time dependencies and size dependencies 
at the equation in the form 
\begin{equation}\label{EqRho}
\frac{a(t)}{\dot a(t)}\left(\frac{\dot \phi(t)}{\phi(t)}+b_0\dot f\right)=\frac{1}{\varrho(y)}\frac{{\rm d}}{{\rm d}y}\omega(y)\varrho(y).
\end{equation}
Next, by  setting  both sides of Eq.(\ref{EqRho}) to the same constant  $-C$ with $C>0$ one gets two separate equations for time dependence
\begin{equation}
\frac{\dot \phi(t)}{\phi(t)}=-C\frac{\dot a(t)}{a(t)}-b_0\dot f
\end{equation}
and bubble size dependence
\begin{equation}
-C\varrho(y)=\frac{{\rm d}}{{\rm d}y}\omega(y)\varrho(y).
\end{equation}  
It allows to get solutions of the form 
\begin{equation}
\begin{split}
\phi(t)&=\phi_0[a(t)]^{-C}\exp(-b_0\dot f t/\tau_\phi),\\
\varrho(y)&=\frac{\varrho_0}{\omega(y)}\exp\left(-C\int^y\frac{{\rm d}y'}{\omega(y')}\right),
\end{split}
\end{equation}
where $\phi_0$, $\tau_\phi$, and $\varrho_0$ are integration constants. 

In such a case one gets
\begin{equation}
\begin{split}
N_b(t)&=N_{0} [a(t)]^{1-C}\exp(-b_0\dot f t/\tau_\phi),\quad N_{0}=\phi_0\int\limits_0^\infty \rho(y){\rm d}y, \\
\left< R_b(t)\right>&=\langle R_{b0}\rangle a(t),\quad  \langle R_{b0}\rangle =\frac{\int_0^\infty y\varrho(y){\rm d}y}{\int_0^\infty\varrho(y){\rm d}y},\quad a(t)\propto t^z.
\end{split}
\end{equation} 
By using experimental data for exponential decaying of the bubble number density (see Ref.\cite{Rest2004}) one can put $C=1$ resulting in $N(t)\propto \exp(-t/\tau_N)$, where $\tau_N=\tau_\phi/b_0\dot f$, the value $\tau_\phi$ is obtained from fitting experimental data.
From obtained solution it follows that the bubble size distribution function has universal character independently on time scale. Moreover the bubble size scale $a(t)\propto t^z$ ($z$ relates to the H\"older exponent, $z=[0,1]$) gives power-law dependence observed at coalescence regime; it characterizes self-similar dynamics of the system and it universal properties, where all measured quantities can be scaled through time scale for bubble size, $a(t)$.   

By taking into account that the swelling is described by the general power law dependence $\Delta V/V\propto (\dot f t)^s$,  one can find $\langle R(t)\rangle\propto  (\dot ft)^{s/3}$, 
where the exponent $s$ defines different kind of governing mechanisms playing a major role in swelling: for vacancy voids one has $s=1$, whereas for totally gas bubbles $s=3/2$ \cite{Rest2004}. By monitoring a change in the slope $z=s/3$ in log-log plot for bubble size versus time one can define a dominating mechanism in bubble growth dynamics.

\section{Results and discussions}

By using material parameters from Tab.\ref{tab1} the modeling procedure was performed at different temperatures and fission rates. The corresponding data analysis was provided at different locations from grain boundary. 

\begin{table}
\caption{Material parameters for $U_3Si_2$ used in modeling\label{tab1}}
\begin{tabular}{c|lcc}
\hline\hline
Parameter&Description&Value&Ref.\\
\hline\hline
$a$ &  Lattice constant  & $7.3314\times 10^{-8}cm$&\cite{RBNR92}\\
$c$ &  Lattice constant  & $3.9001\times 10^{-8}cm$&\cite{RBNR92}\\
$\Omega_0$ & Atomic volume of $U$ in $U_3Si_2$ crystal structure&$3.629\times 10^{-23}cm^{3}$&\cite{AABM2020}\\
$\Omega_v$ & Vacancy volume & $4.09\times10^{-23} cm^3$&\cite{Kogai97}\\
$\Omega_g$ & van der Waals gas atom volume & $8.5\times10^{-23} cm^3$&\cite{Barani2019}\\
$r_{iv}$&Defect recombination radius& $8.25\times 10^{-7}cm$&\cite{Rest2000}\\
%$\rho_{Xe}$& Solid state density of Xe& $3.64g/cm^{3}$&\\
%$A_{Xe}$& atomic mass of Xe& $131.293g/mol$&\\
$b_v$ & Burgers vector&$\frac{1}{2}a<110>$&\\
$G$ &Number of survived Frenkel pairs per fission& $G\in[1000,\textbf{3000}]$& \cite{HSB2020}\\
$\dot f$&Fission rate& $(1.7-10.0)\times 10^{14} s^{-1} cm^{-3}$&\cite{Rest2000}\\ 
%\hline
%$f_n$& &$10^{-4}$&\cite{Vesch2000}\\
$f_n$ & Bubble nucleation factor &$10^{-2}$&\cite{Report2017}\\
%\hline
%%%%%%%%%%%%%%%%%%%%%%%%%%%%%%%%%
$\beta$& Gas atoms produced per fission& 0.25& \cite{Olander76}\\
$r_g$ & Radius of diffusion gas atom & $0.216\times 10^{-7}cm$& \cite{Rest1994}\\
%\hline
$b_0$ & Bubble resolution volume& $1\times 10^{-19}cm^{3}$& \cite{Report2017}\\
%\hline
$\dot f_0$& Effective fission rate&$1.25\times 10^{14}s^{-1}cm^{-3}$& \cite{Rest2000}\\
$\eta_0$& Effective viscosity&$2\times 10^{7}\dot f_0~ poise$& \cite{Rest2000}\\
$\alpha_0$ & Proportionality constant& $5\times 10^{-10}\dot f_0$ $s^{-1}cm^{-3}$& \cite{Rest2000}\\
%$\gamma$ & Surface tension ($U_3Si_2$)& $0.7~J/m^2$& \cite{Finlay04}\\
%$\gamma$ & Surface tension ($U_3Si_2$)& $1.16~J/m^2$& \cite{MM2016}\\
%$\gamma_{\{100\}}$,$\gamma_{\{001\}}$ & Surface tension ($U_3Si_2$)& $1.48~J/m^2$, $1.43~J/m^2$& \cite{MGA2017}\\
%$\gamma$&\textbf{ Mean surface tension ($U_3Si_2$)}& $1.32\times 10^{-4} J/cm^2$&\cite{MM2016}\\
%$E_i^f$& Interstitial formation energy ($UMo$)&$0.8 eV$&\cite{HSB2020}\\
%$E_v^f$& Vacancy formation energy ($U_3Si_2$)&$1.69 eV$&\cite{ALB2019}\\
%$E_g^f$& Gas atom (Xe) formation energy in $U$ matrix&$4.93 eV$&\cite{ALB2019}\\ 
%$\beta_v$ & van der Waals constant& $0.857\times 10^{-22}cm^3$&\cite{Rest2000} \\
%$k_v^m=k_g^m$&  Fitting parameters& $5.28\times 10^{5}J/cm^3$&\cite{AABM2020}\\
%$k_v^b=k_g^b$&  Fitting parameters& $8.0\times 10^{4}J/cm^3$&\cite{AABM2020}\\
%$c_g^{b,eq}$& Eq. gas (Xe) atom concentration in bubbles& 0.3924&\cite{AABM2020}\\
%$c_v^{b,eq}$& Eq. vacancy concentration in bubbles& 0.6076&\cite{AABM2020}\\
%$E_i^S$& Interstitial-sink interaction energy & $2.0 eV$&\cite{HSB2020}\\
%$E_v^S$& Vacancy-sink interaction energy & $-2.0 eV$&\cite{HSB2020}\\
 %$E_g^S$& Gas atom-sink interaction energy &$-2.0eV$&\cite{HSB2020}\\
%\hline 
%$D_g^{RED}$& Proportionality constant ($U0_2$) &$3.5\times 10^{-30} cm^5$&\cite{Rest2000}\\
$D_{g}^{RED}$& Proportionality constant  &$1.0\times 10^{-6} cm^5$& \cite{Report2017}\\
%\hline
%$D_g^0$& Gas diffusion coefficient ($U_3Si_2$)& $5.91\times 10^{-2}\exp(-4.41\times 10^{-19}J/T) cm^{2}/s$&\cite{Barani2019}\\
$D_g^0$& Gas atoms diffusion coefficient & $7.73\times 10^{-3}\exp(-1.68eV/T) cm^{2}/s$&\cite{Report2017}\\
%\hline
%$D_v$ & Intra-granular vacancy diffusivity ($U_3Si_2$)&$3.35\times10^{-2}\exp(-4.63\times 10^{-19} J/T) cm^{2}/s$&\cite{ALB2019}\\
$D_v$ & Vacancy diffusivity  &$7.53\times10^{-2}\exp(-2.13eV/T) cm^{2}/s$&\cite{Report2017}\\
$D_i$ & Interstitial diffusivity &$1.05\times10^{-1}\exp(-1.68eV/T) cm^{2}/s$&\cite{Report2017}\\
%$D_v^{gb}$ & Inter-granular vacancy diffusivity &$10^6\times D_v cm^{2}/s$&\cite{Barani2019}\\
%$\omega$ & The grain boundary width& $3\times 10^{-8}cm$&\cite{Cheniour2020}\\
%$\overline{\gamma}$& Grain boundary energy&$0.83 J/m^2$ at $673 K$&\cite{Cheniour2020}\\
%$\delta_{gb}$& Thickness of diffusion layer in grain boundaries& $5\times 10^{-8}cm$& \cite{Kogai97}\\
%$\sigma_0$ & Interfacial energy ($U_3Si_2$)& $1.7\times 10^{-4} J/cm^2$& \cite{AABM2020}\\
%$\delta_0$ & Interface thickness ($UMo$) & $3.8\times10^{-6} cm$& \cite{HSB2020}\\
%$F_0$& Constant free energy barrier& $6.84\times 10^8J/m^3$&\cite{AABM2020}\\
%$\gamma_{ij}$& Grain interface energy parameter & $0.6749$&\cite{AABM2020}\\
%$\varkappa$&Gradient energy coefficient&$1.92\times 10^{-8}J/m$&\cite{AABM2020}\\
%$V_m$& Molar volume of $U$& $1.2495\times10^{-5}m^3/mol$&\\
%$L_\eta$& Order parameter mobility &$1.59\times 10^{-4}cm^3/(Js)$&\cite{AABM2020}\\
$\nu$&Poisson ratio & 0.31&\cite{Rest2004}\\
%$\langle K\rangle$& Bulk modulus (U)& 101 GPa&\\
$\lambda_{GB}$& Grain size& $20\times10^{-4}cm$&\cite{HSB2020}\\
%$T$&(Preparation)Operating temperature&$(0.9T_m),\quad  (800-1200)K$&\cite{MGAMY2018}\\
%$\Delta V/V$& Swelling in LWR&$(10-12)\%$&\cite{Sh65}\\
%$F.D.$& Fision density&  $2.5-5\times 10^{21} cm^{-3}$&\cite{Finlay04}\\
%\hline
%$c_{i,v}$ & Interstitial/vacancy concentration & at.fract.&\\
%$c_g$ & Gas atoms concentration in fuel&$cm^{-3}$&\\
%$R_b$   & Cavity/bubble radius&$(2-7)nm\to(0.1-0.15)\mu m$&\cite{Barani2019,Finlay04}\\
%$m_b$, $n_{vb}$ & Number of atoms/vacancies in each bubble& &\\
%$N_b$ & Cavity/bubble concentration&$(2-9)\times10^{13}cm^{-3}$&\cite{Finlay04}\\
\hline\hline
\end{tabular}
\end{table}

\begin{figure}
\centering
a)\includegraphics[width=0.4\textwidth]{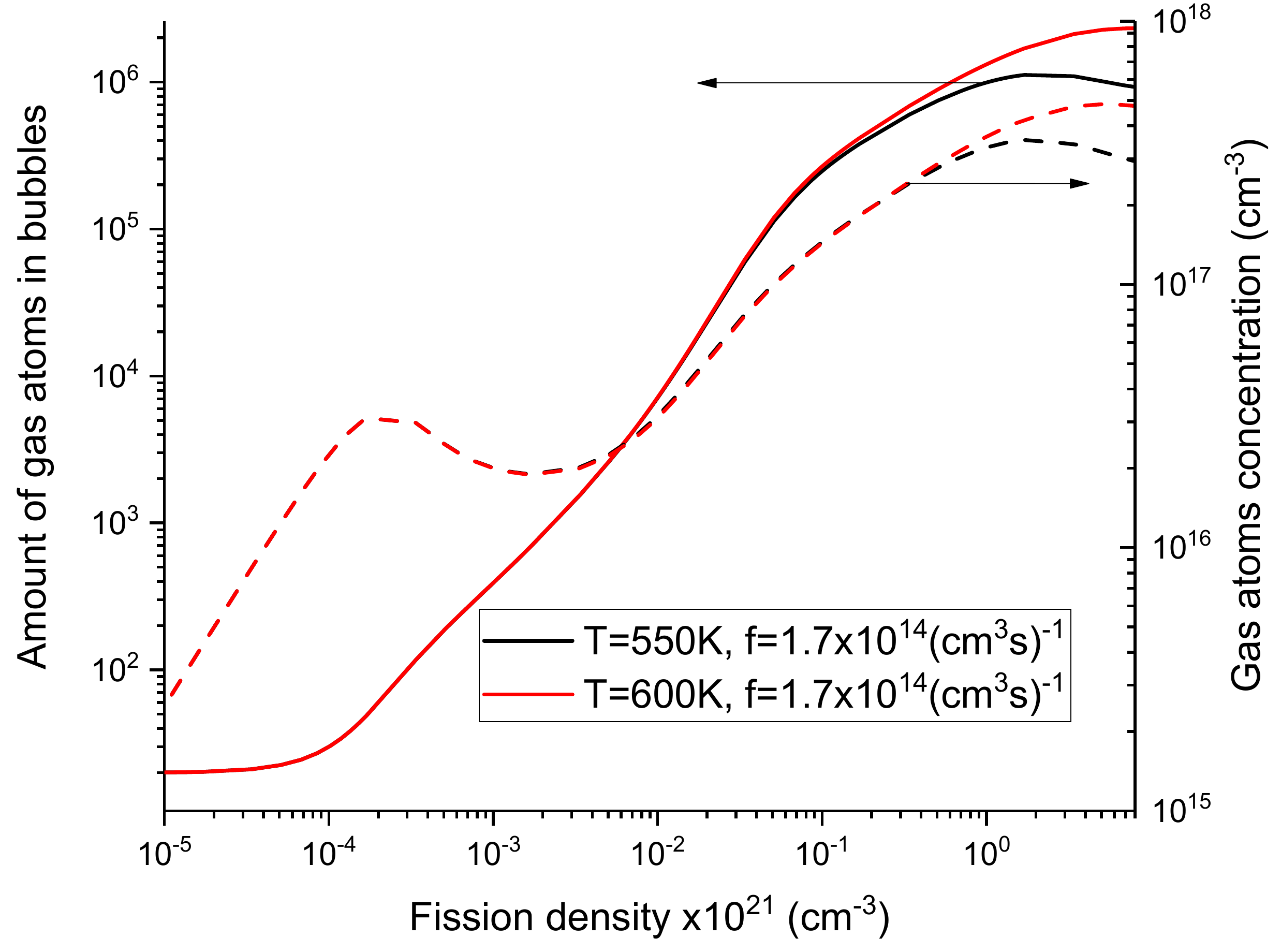}
b)\includegraphics[width=0.4\textwidth]{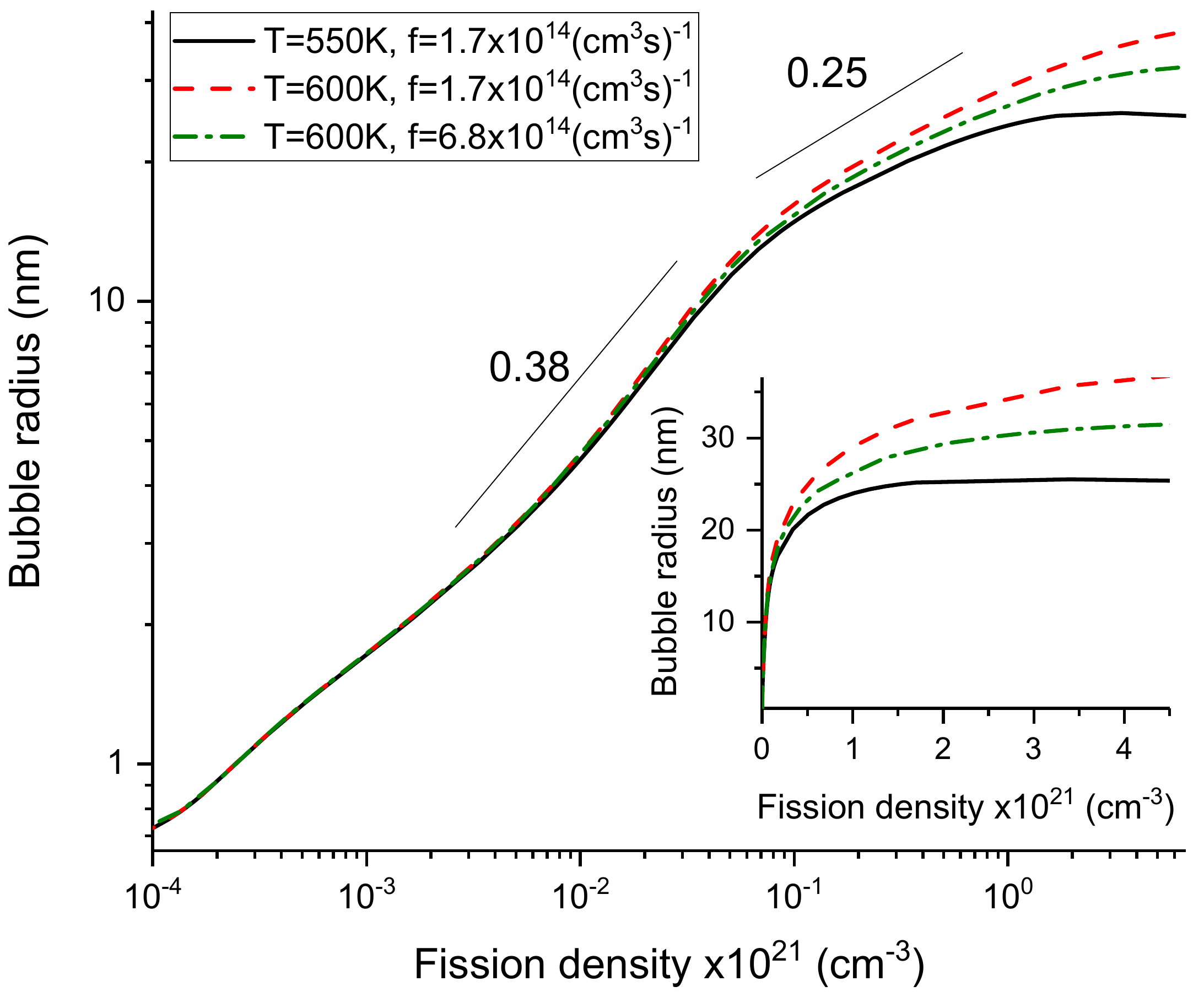}\\
c)\includegraphics[width=0.4\textwidth]{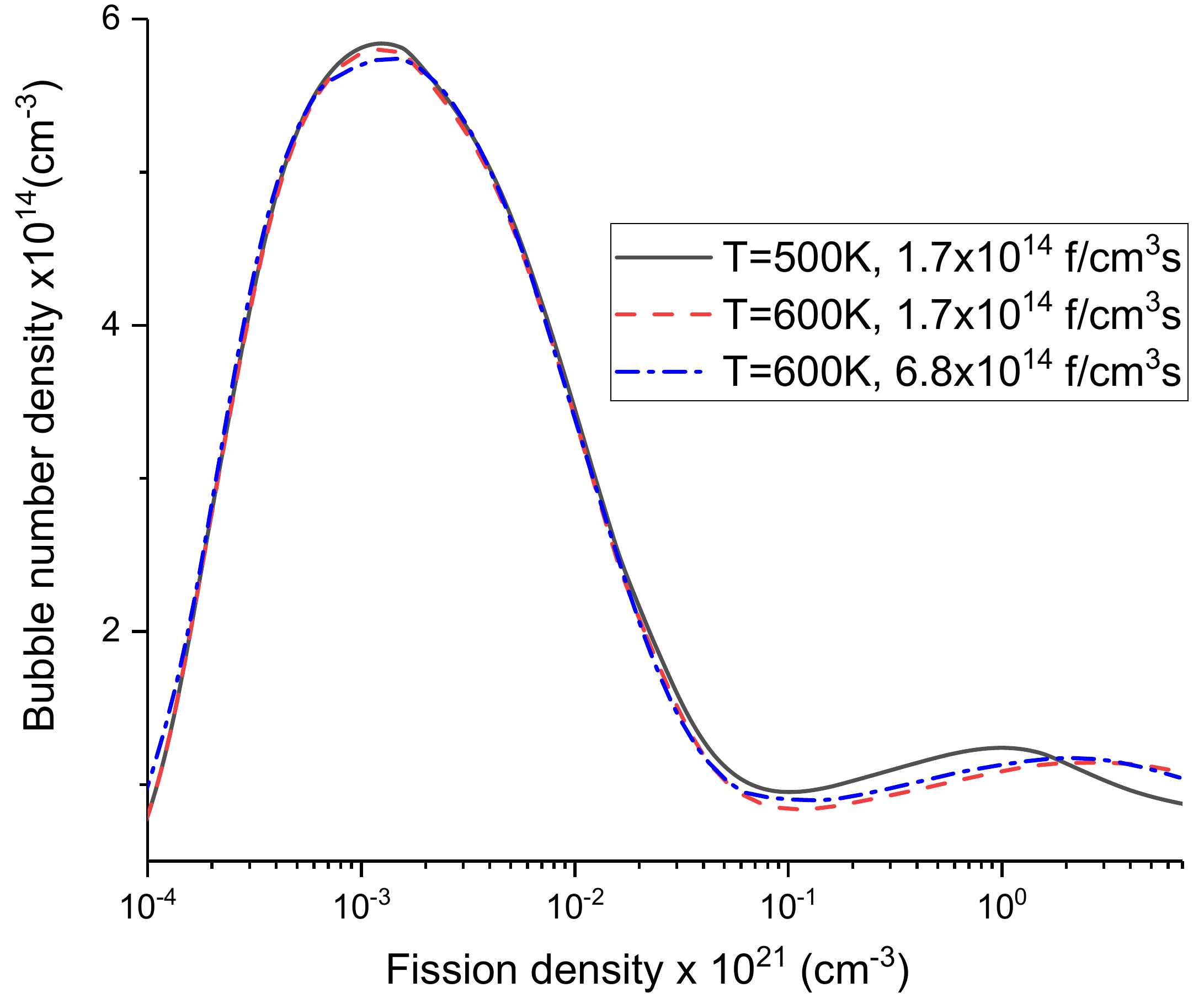}
d)\includegraphics[width=0.4\textwidth]{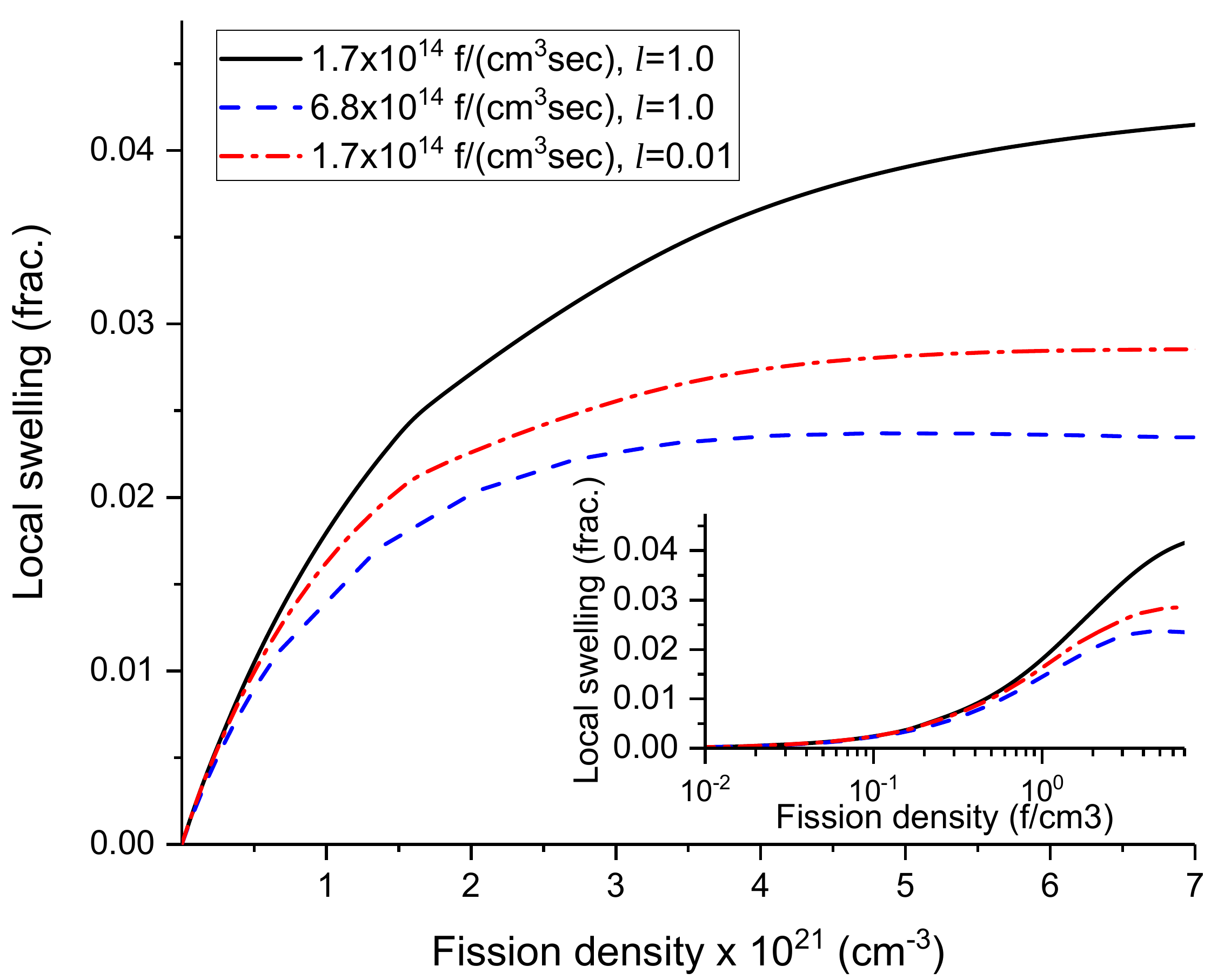}\\
e)\includegraphics[width=0.4\textwidth]{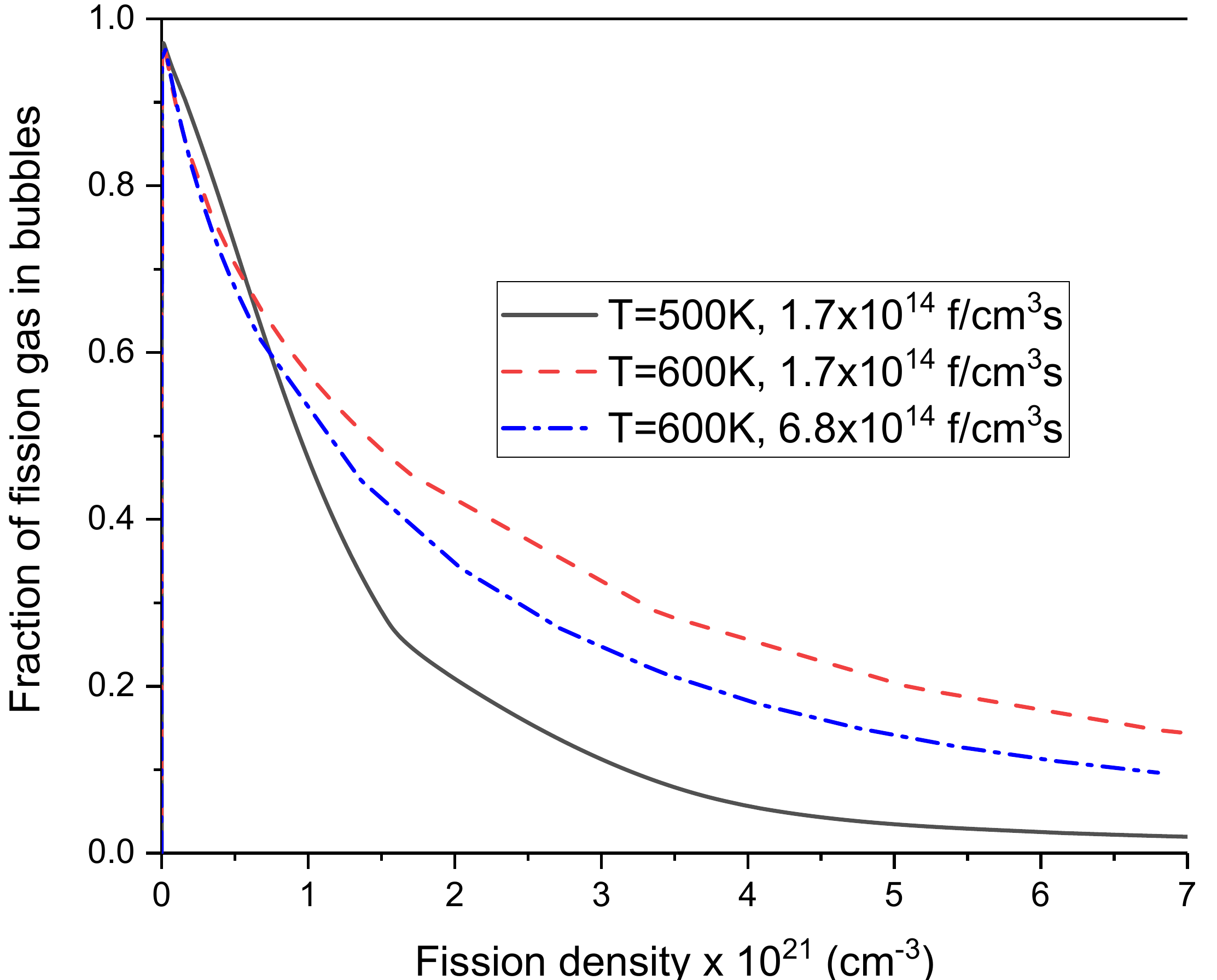}
f)\includegraphics[width=0.4\textwidth]{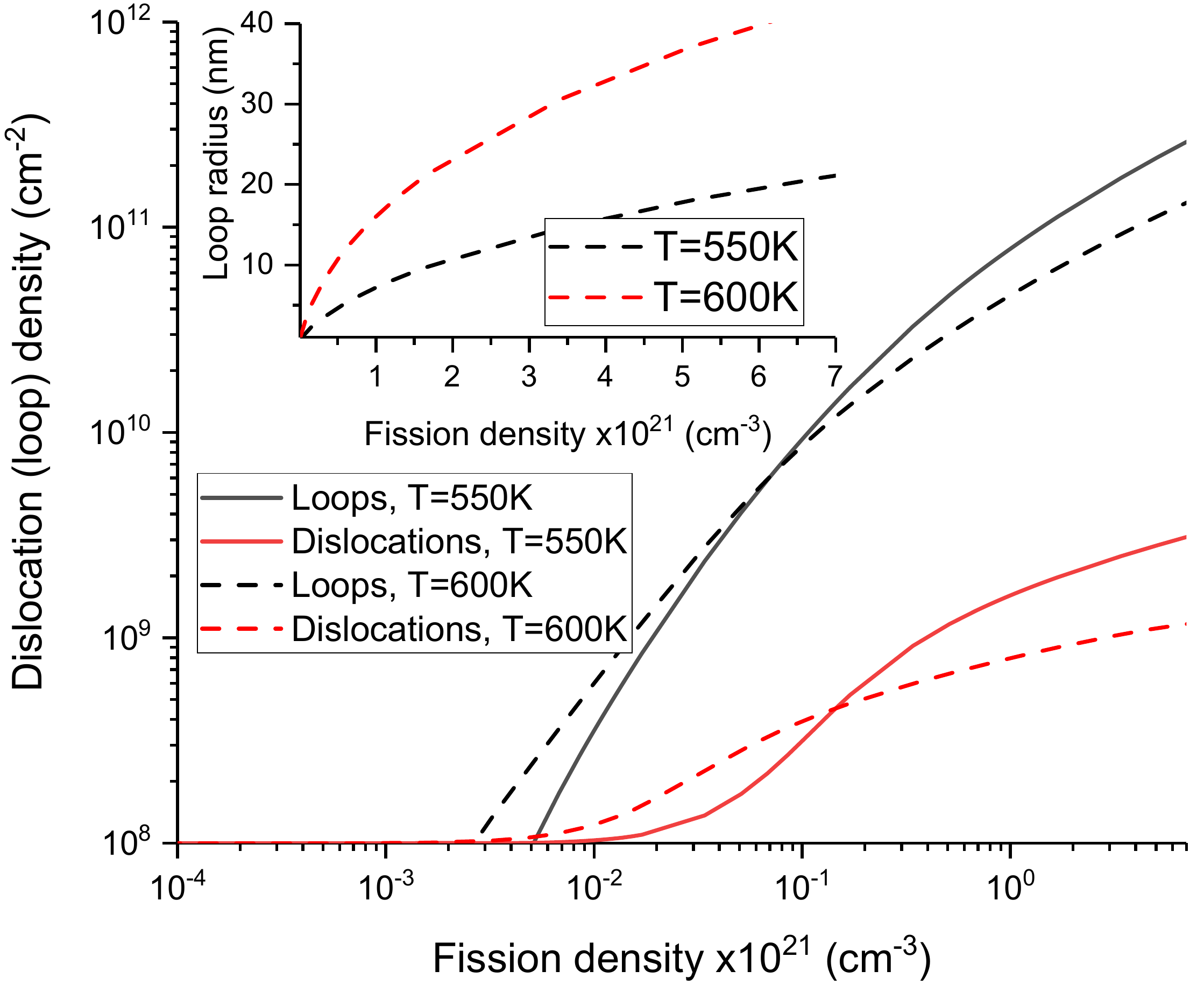}
\caption{Fission density dependencies: (a) number of gas atoms in bubbles and gas atoms concentration, (b) bubble radius, (c) bubble number density, (d) local swelling, (e) fraction of fission gas in bubbles, and (f) loop and dislocation densities  at different parameters indicated in legend. Plots a,b,c,e,f correspond to grain center domain. Plot d obtained at $T=600K$. Plots a and f  correspond to  $\dot f=1.7\times 10^{14}f/cm^3s$\label{fig_protocols}}
\end{figure}

Typical protocols giving main information about gas bubble growth are shown in Fig.\ref{fig_protocols}.  From   Fig.\ref{fig_protocols}a,b one finds  a crossover of dynamical regimes: incubation at small fission density, power law asymptotic at a growth stage and quasistationary stage of slow dynamics at elevated fission densities.  An accumulation of gas atoms in matrix at small fission densities results in formation of gas bubbles, the number of gas atoms in bubbles starts to grow after incubation time. Here the concentration of gas atoms manifests a decaying dynamics, whilst number of gas atoms in bubbles grows crucially (cf. solid and dashed curves in Fig.\ref{fig_protocols}a). With further fission gas atoms concentration increases and bubble size grows up  by accumulating more gas atoms inside (Fig.\ref{fig_protocols}b). A decaying dynamics  of both gas concentration and atoms number in bubble at elevated fission densities and low temperatures results in small decrease in the bubble size due to low diffusivity of gas atoms toward bubbles comparing to gas atoms resolution.  From the other hand an increase in fission density decreases bubble size due to growth of the resolution rate. We indicate here a slope change of bubble size dependencies by putting the values of $z=s/3$. It follows that at the stage of bubble growth one gets $s\approx 1.14$ that is close to classical Lifshitz-Slyozov asymptotic law, whereas at large fission densities one gets $s\approx 0.75-0.8$ responsible for slow dynamics of large bubbles size. The mean bubble size increases up to 40nm within the temperature range 500-600K. Protocols for bubble number density (Fig.\ref{fig_protocols}c) illustrate exponential decaying dynamics at the stage of bubble size growth. It relates to a coalescence scenario when an interaction of bubbles leads to formation of large bubbles by absorbing small ones. Here due to an increase in a critical size of bubbles small ones became unstable whilst overcritical bubbles grow slowly in quasistationary manner. A stage of the number density decrease  with bubble size growth is well observed on protocols for swelling starting from  the  so-called ``knee point''. At this point due to the  mass coalescence the swelling curve manifests an increase in a slope (see Fig.\ref{fig_protocols}d and insertion there). As it is seen from swelling dependencies with a further  fission density growth one gets slowing down dynamics caused by slow dynamics of bubble size. 
Decaying behavior of the fission gas fraction in bubbles is shown in Fig.\ref{fig_protocols}e. It follows that at small fission density the corresponding value increases rapidly, whereas at increased fission density this fraction decays down to 10-20\% depending on temperature and fission rate.  
In our modeling routine we consider dynamics of point defect sinks, the corresponding protocols are shown in Fig.\ref{fig_protocols}f. Here one can observe  an increase in both  loop density and network dislocation density  with different stages, where loop density is higher by two orders comparing to network dislocations at elevated temperatures. The  loop radius increases with temperature and attains values  up to 40nm at elevated fission densities.  

\begin{figure}[!t]
\centering
\includegraphics[width=0.4\textwidth]{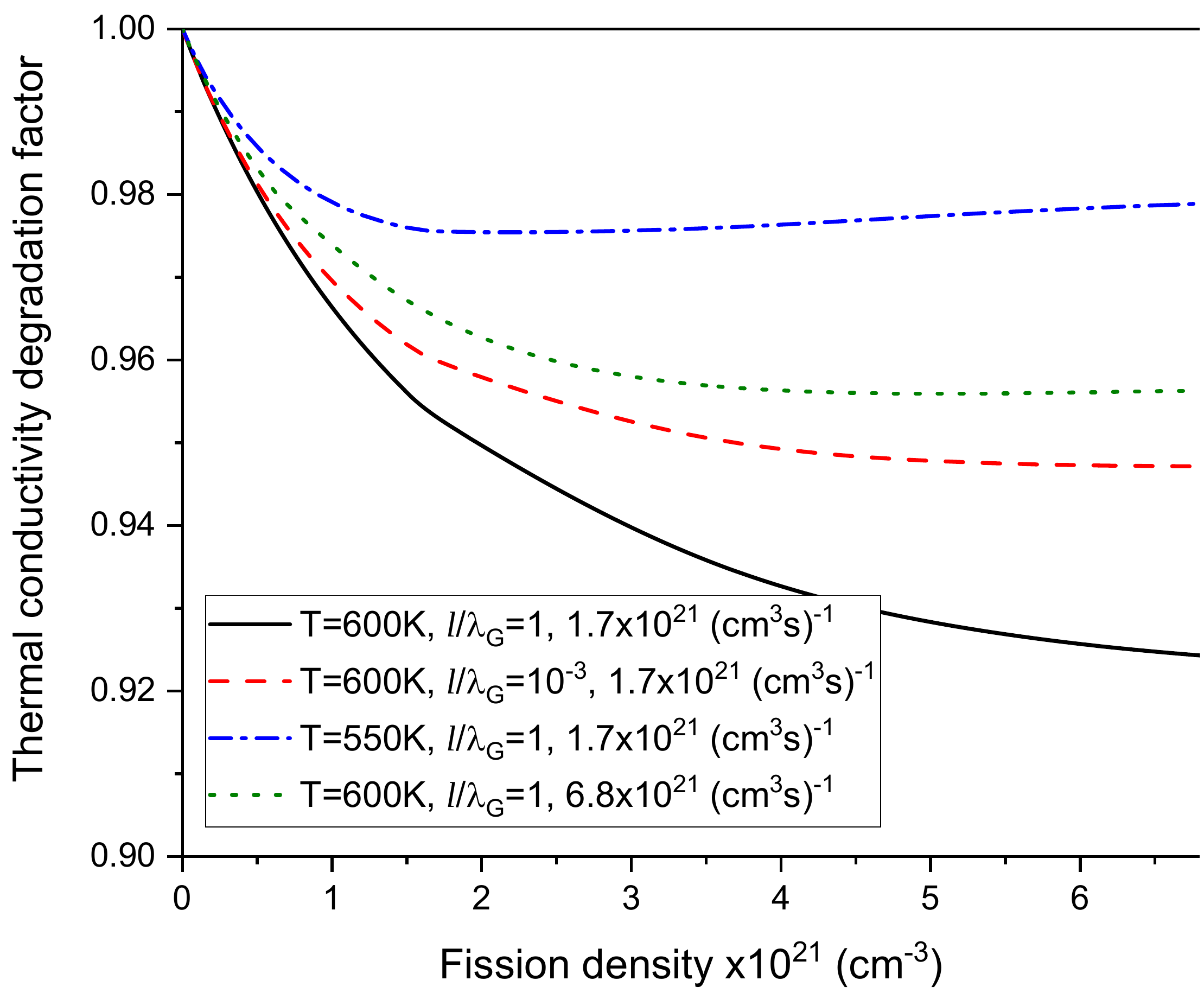}
\caption{Dependence of the thermal conductivity degradation factor at  different temperatures, fission rate and location in the grain\label{kappaFd}}
\end{figure}

Obtained data can be used to estimate reduction of thermal conductivity due to bubbles inhibiting heat transfer through $U_3Si_2$. 
A porosity impact on thermal conductivity we compute by using the Maxwell-Eucken
equation, which gives a  multiplier to the
thermal conductivity $\varkappa=(1-\sigma)/(1+0.9\sigma)$, where $\varkappa$ is the thermal conductivity degradation factor, $\sigma=N_bV_b$ stands for volume fraction of bubbles  \cite{Tonks2016}. 
Dependence of  this factor on fission density is shown in Fig.\ref{kappaFd}. One can observe  a decrease of thermal conductivity  from several percents down to 10\% at elevated fission densities ($6.8\times 10^{21} (cm^3s)^{-1}$) and increased temperatures.   

\begin{figure}[!t]
\centering
\includegraphics[width=0.4\textwidth]{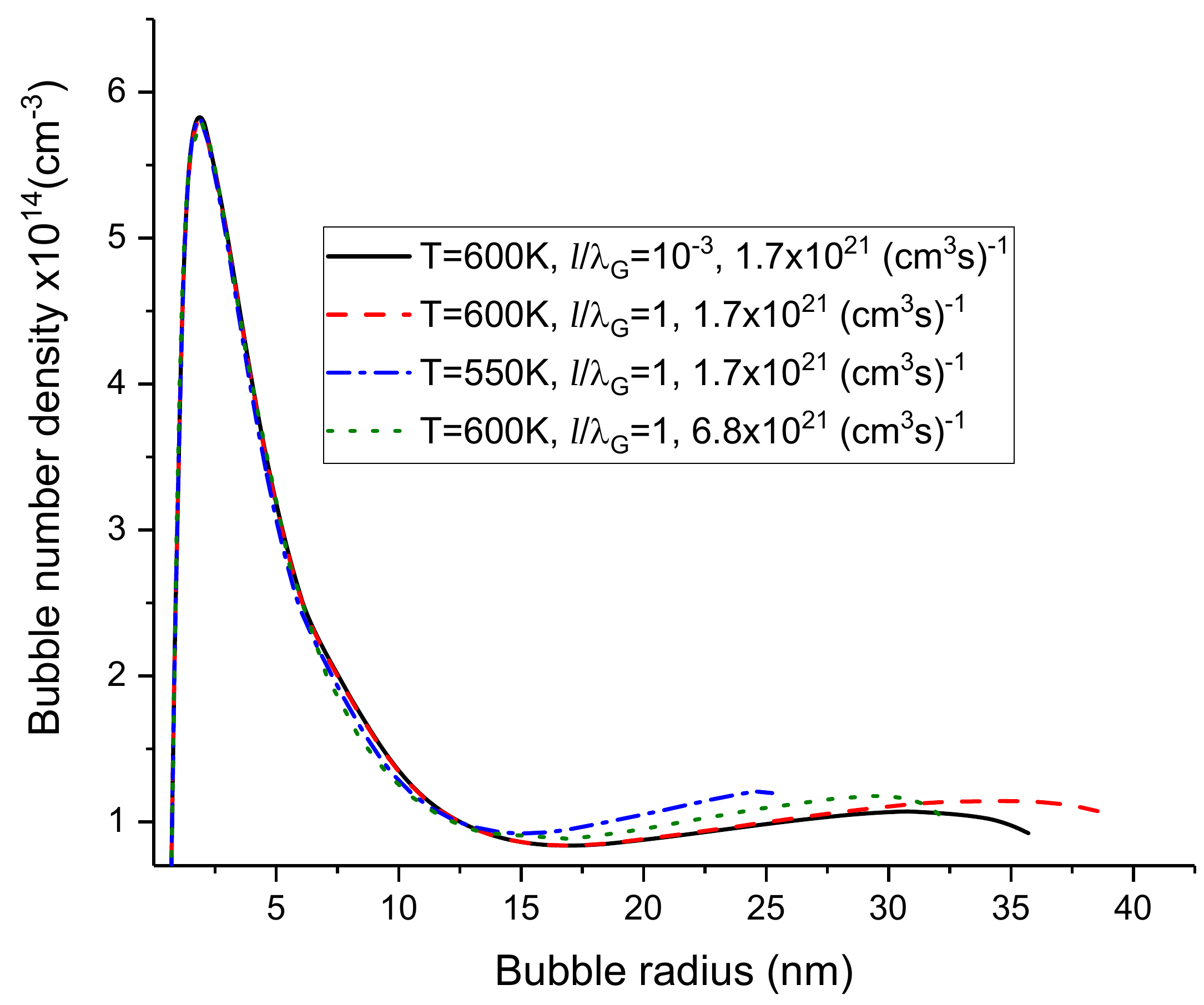}
\caption{Dependence of the number density on bubble size  at  different temperatures, fission rate and location in the grain\label{Nd_Rb}}
\end{figure}

By analyzing time dependencies of both bubble radius and the corresponding number density one finds that at elevated fission densities where the bubble size increases up to 20-30nm  the number density manifests second peak. It means that the amount of large bubbles increases in grains. By plotting dependencies of   number density versus bubble radius one can observe two peaks (see Fig.\ref{Nd_Rb}): the first one relates to small bubbles of the size around 2nm whilst the second one corresponds to large bubbles. Obtained two-peak dependence has the same character as was shown in Ref.\cite{MKGAYMHY2017}. It follows that small bubbles are distributed along the grain with the same density, whereas the number density of large bubbles is lower in the grain boundary vicinity comparing to the grain center.A temperature decrease results in increasing amount of intermediate size bubbles (around 25nm) but prevents formation large ones. At the same time an increase in the fission rate increases the amount of intermediate size bubbles but  does not support formation of large ones.   

\begin{figure}[!t]
\centering
a)\includegraphics[width=0.4\textwidth]{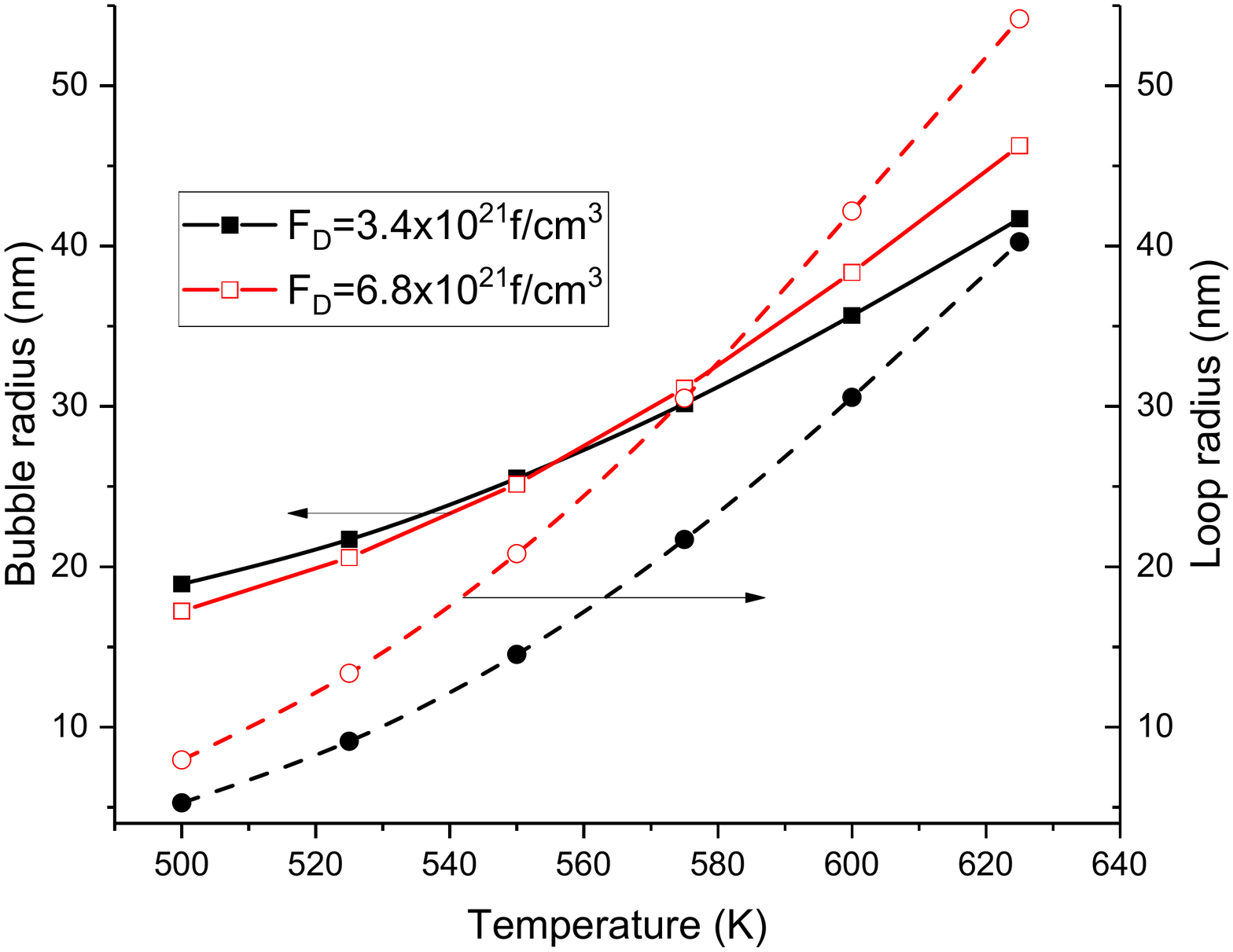}
b)\includegraphics[width=0.4\textwidth]{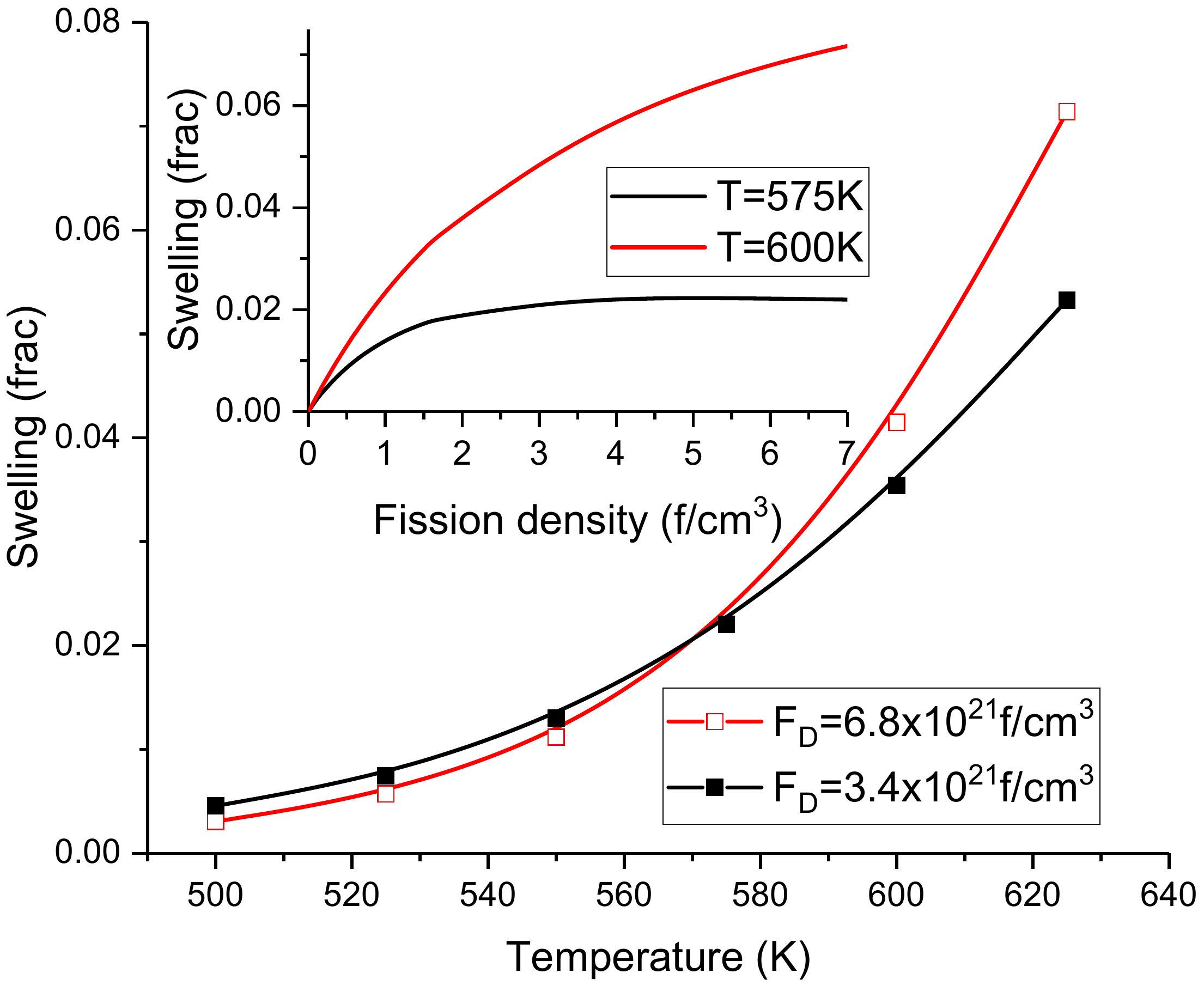}\\
c)\includegraphics[width=0.4\textwidth]{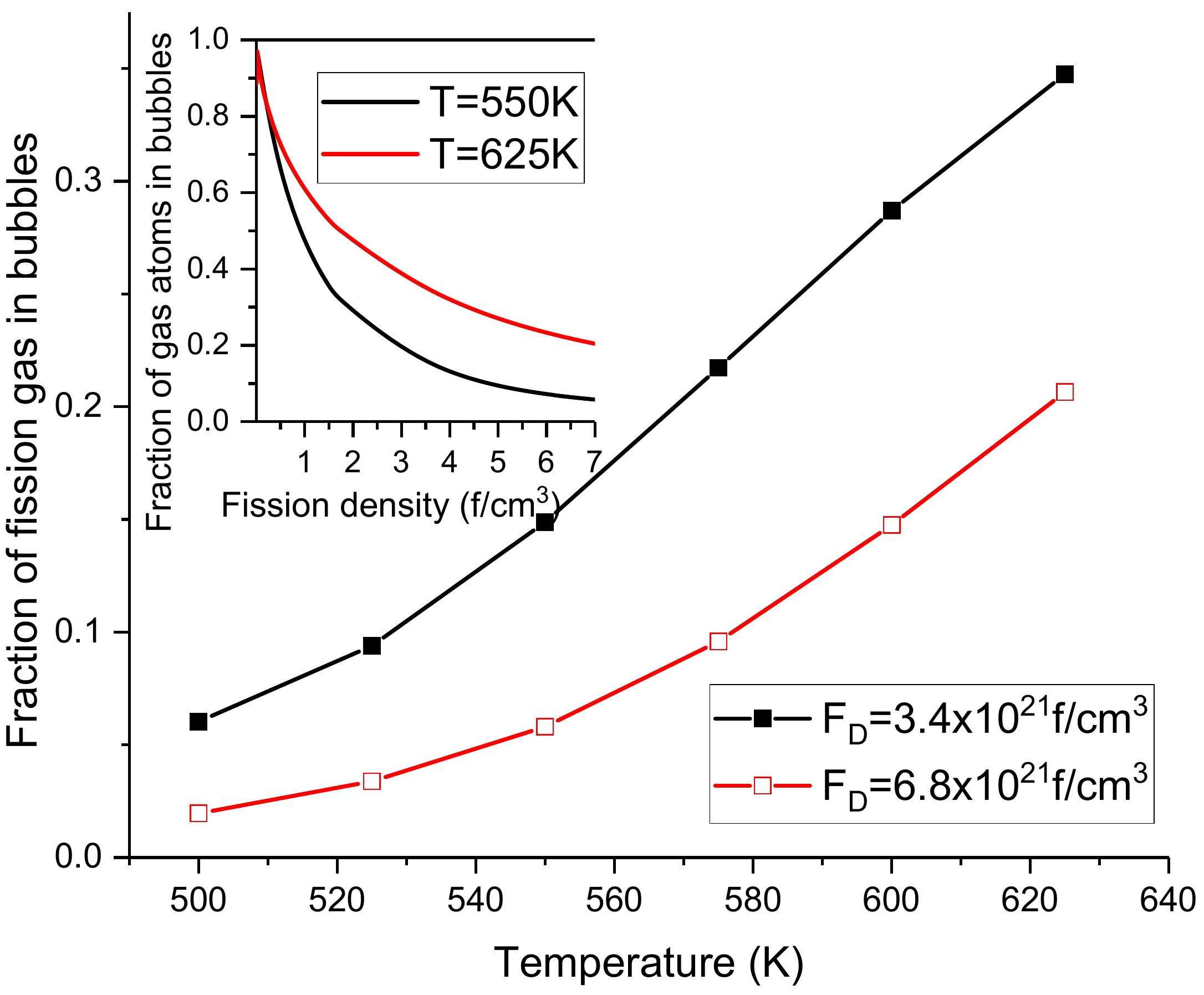}
d)\includegraphics[width=0.4\textwidth]{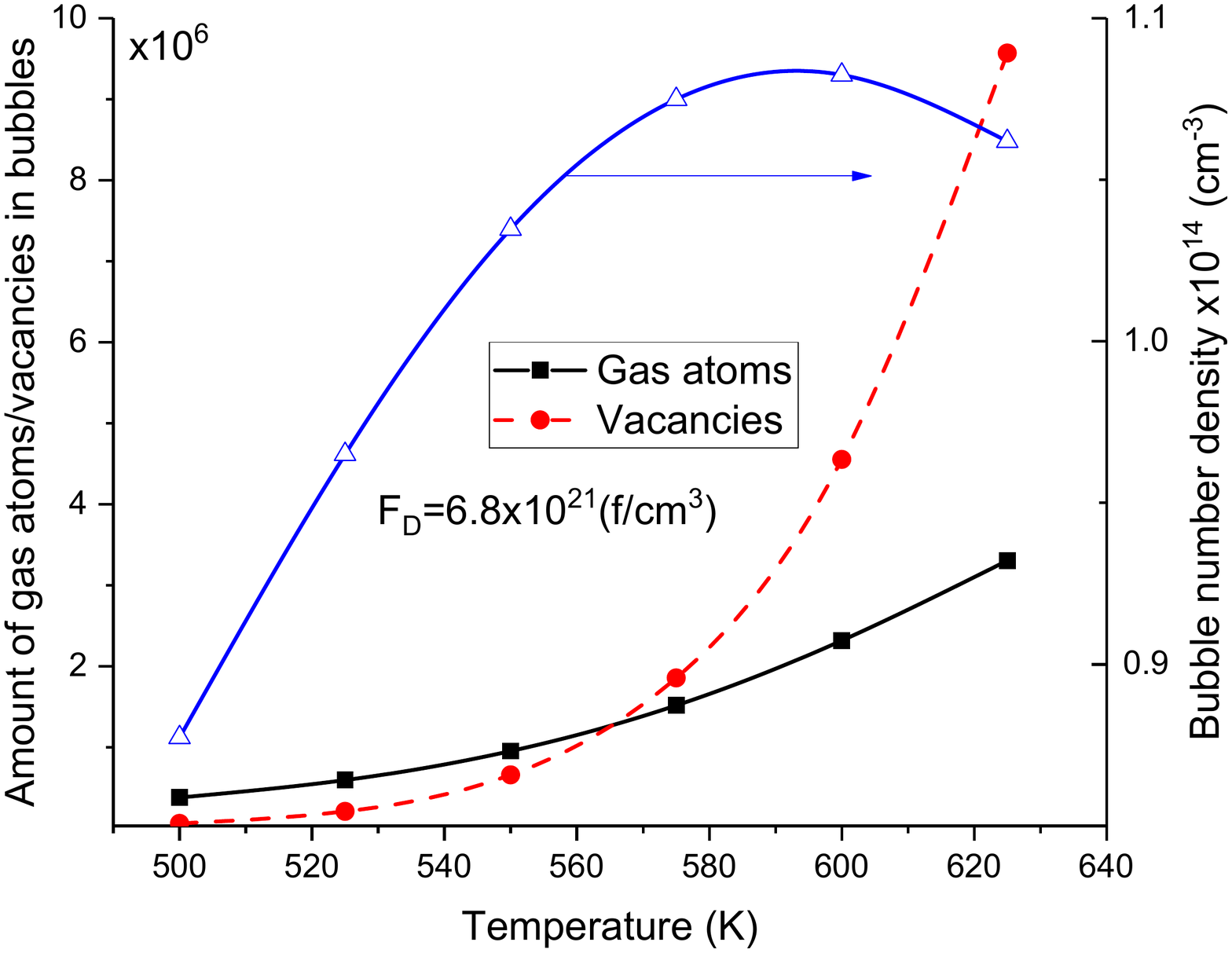}\\
e)\includegraphics[width=0.4\textwidth]{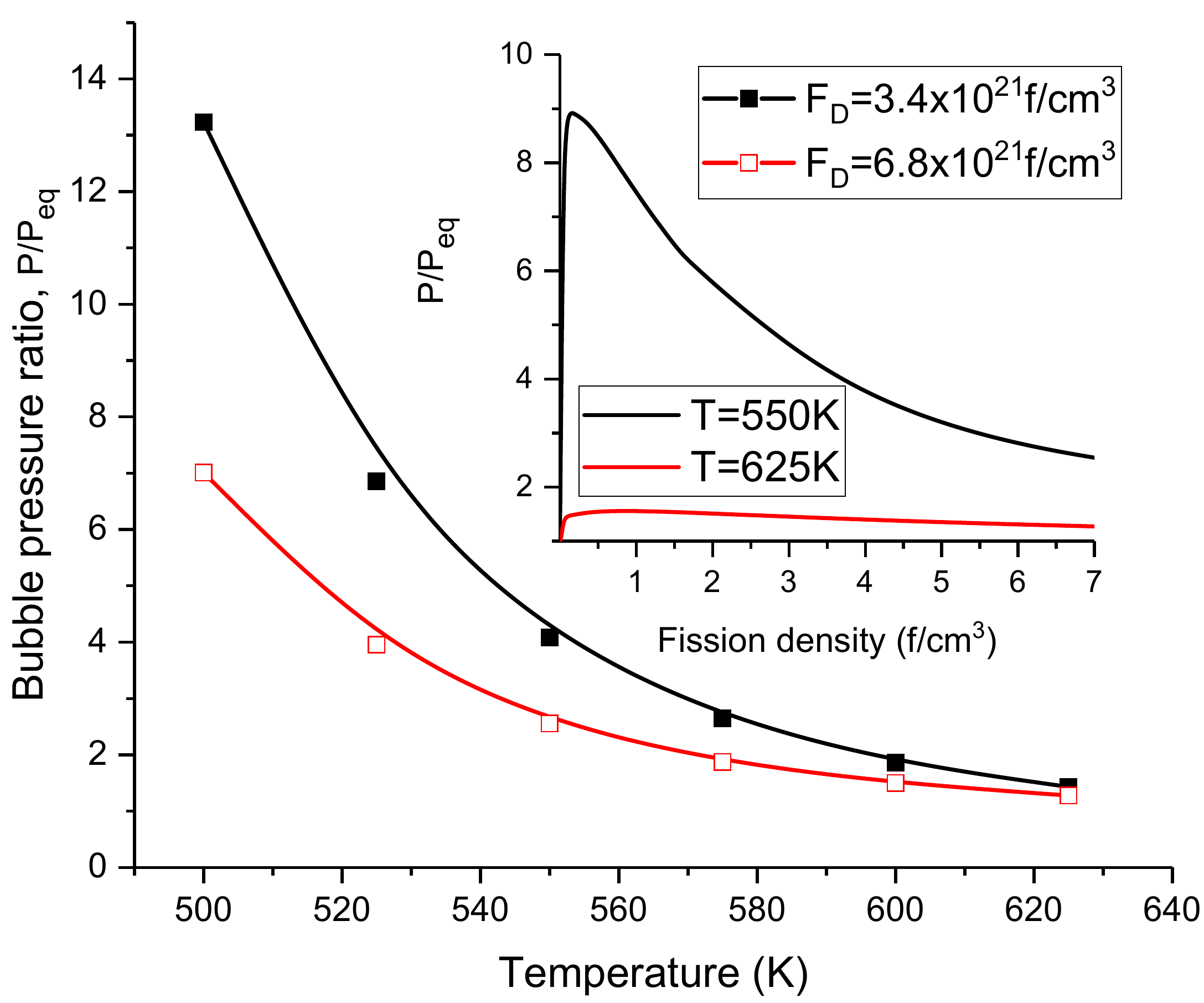}
f)\includegraphics[width=0.4\textwidth]{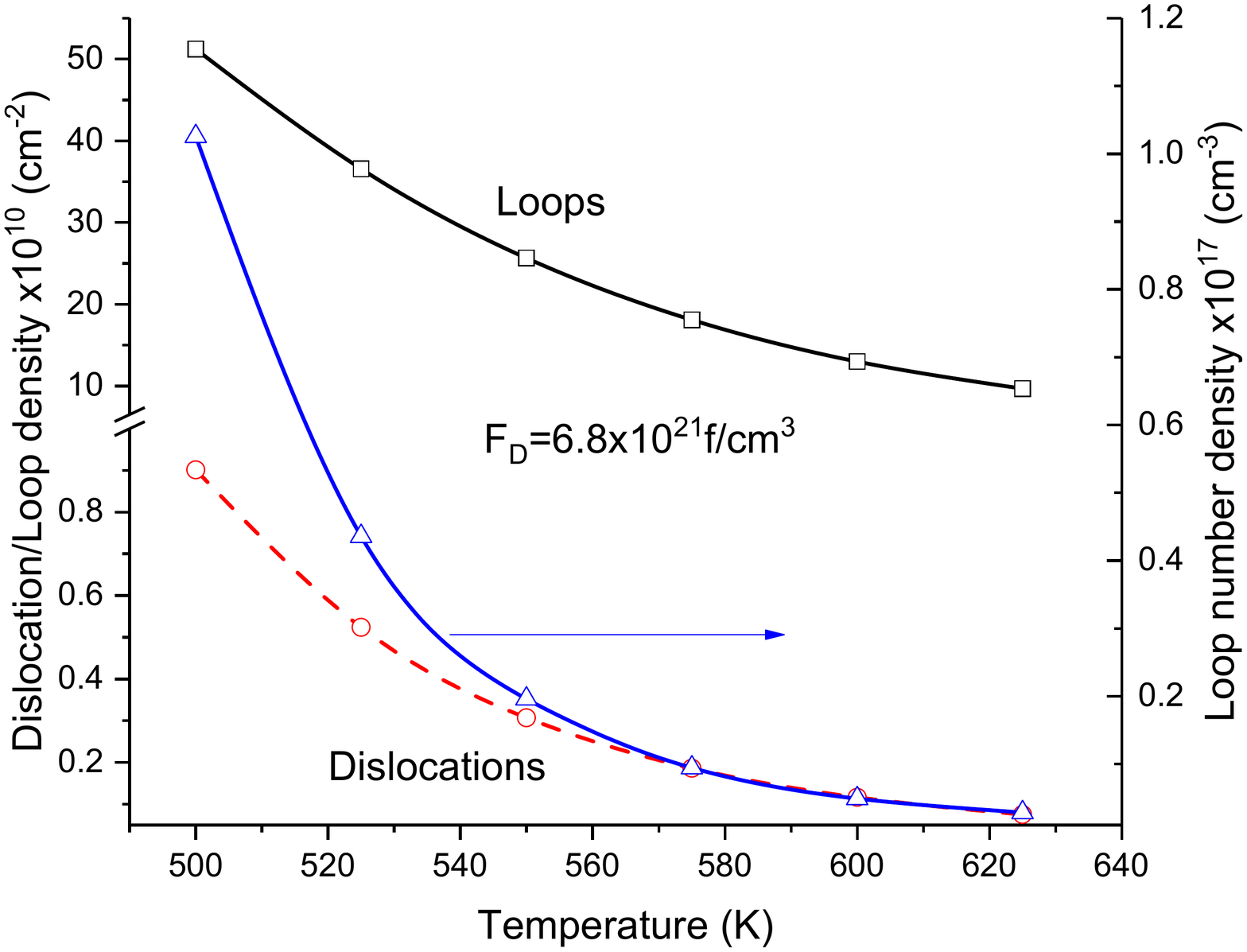}
\caption{Temperature dependencies at the grain center: plots a, b, c relate to  bubble and loop radii, swelling and frcation of fission gas in bubbles; plot d gives values of amount of gas atoms and vacancies in bubbles togethre with bubble number density at fixed fission density; plots e, f correspond to bubble pressure dependncies and loop and dislocation densities with loop number densities. All data are obtained at the fission rate $1.7\times 10^{14}f/(cm^3 s)$\label{Tdep1}}
\end{figure}

Considering  temperature effect one finds that with the temperature increase the net flux of point defects and gas atoms grows up to bubbles due to increasing diffusivity of defects and gas atoms.  It leads to a growth of bubbles and dislocation loops whose  sizes become comparable by mean values  varying from 20nm up to 45nm for bubbles and from 10nm up to 55nm for loops with temperature (see Fig.\ref{Tdep1}a). The fractional swelling increases with temperature (see Fig.\ref{Tdep1}b). 
By comparing temperature dependencies of bubble size and swelling at different fission densities  one finds that at low temperatures  both bubble size and swelling have lower values at elevated fission densities than at higher temperatures (cf.filled ad empty square markers in Fig.\ref{Tdep1}a,b). By comparing swelling dynamics at different temperatures (see insertion in Fig.\ref{Tdep1}b) one arrives at the conclusion that fast growth at elevated temperatures is caused by increasing of both the bubble size and their number density as shown in Fig.\ref{fig_protocols}b,c. Obtained values of swelling  at discussed temperature interval relates well to results shown in Ref.\cite{MKGAYMHY2017}.
  In Fig.\ref{Tdep1}c,d we plot temperature dependencies of fraction of fission gas in bubbles computed as $m_bN_b/\beta\dot f t$ and  amount of gas atoms and vacancies in a bubbles with number density of bubbles. A decaying dynamics of this fraction (see insertion in Fig.\ref{Tdep1}c) illustrates that with fission density increase the fraction of gas in bubbles goes down in the same manner  as was discussed in Ref.\cite{Rest2004}. This fraction increases with temperature due to an increase in both number of bubbles and number of vacancies and gas atoms in bubble that is well seen from Fig.\ref{Tdep1}d.  
Here one concludes that at low temperatures these bubbles contain more gas atoms than vacancies. With the temperature increase fast migration of vacancies promotes their fast diffusion to bubbles resulting in appearance of larger amount of vacancies comparing to gas atoms. Such amount of vacancies decreases the pressure in gas bubbles (see Fig.\ref{Tdep1}c), but the bubble size growth results in increase in swelling  with temperature at studied temperature range. 
Temperature dependencies of  pressure ratio $p/p_{eq}$ and the corresponding protocols are  shown in Fig.\ref{Tdep1}e. As far as $p/p_{eq}>1$ it follows that there is some deficit of vacancies in bubbles resulting in accumulation of vacancies from matrix at low temperatures. At elevated temperatures this ratio goes to 1 and takes lower values resulting in emission of vacancies from bubbles. From protocols shown in insertion in Fig.\ref{Tdep1}e it is well seen fast increase of pressure in bubbles and its further decay with fission density. It is caused by  a decrease of the equilibrium pressure ($p_{eq}\propto R_b^{-1}$) and increase of the number of vacancies in bubbles comparing to number of gas atoms ($p\propto m_b/n_{vb}$).
  By considering temperature dependencies of loop characteristics (see Fig.\ref{Tdep1}f) one gets an increase of the mean loop radius accompanied by a decrease in loop density. This effect is caused by a decrease of the loops number density (see insertion in Fig.\ref{Tdep1}f).

\begin{figure}[!t]
\centering
a)\includegraphics[width=0.4\textwidth]{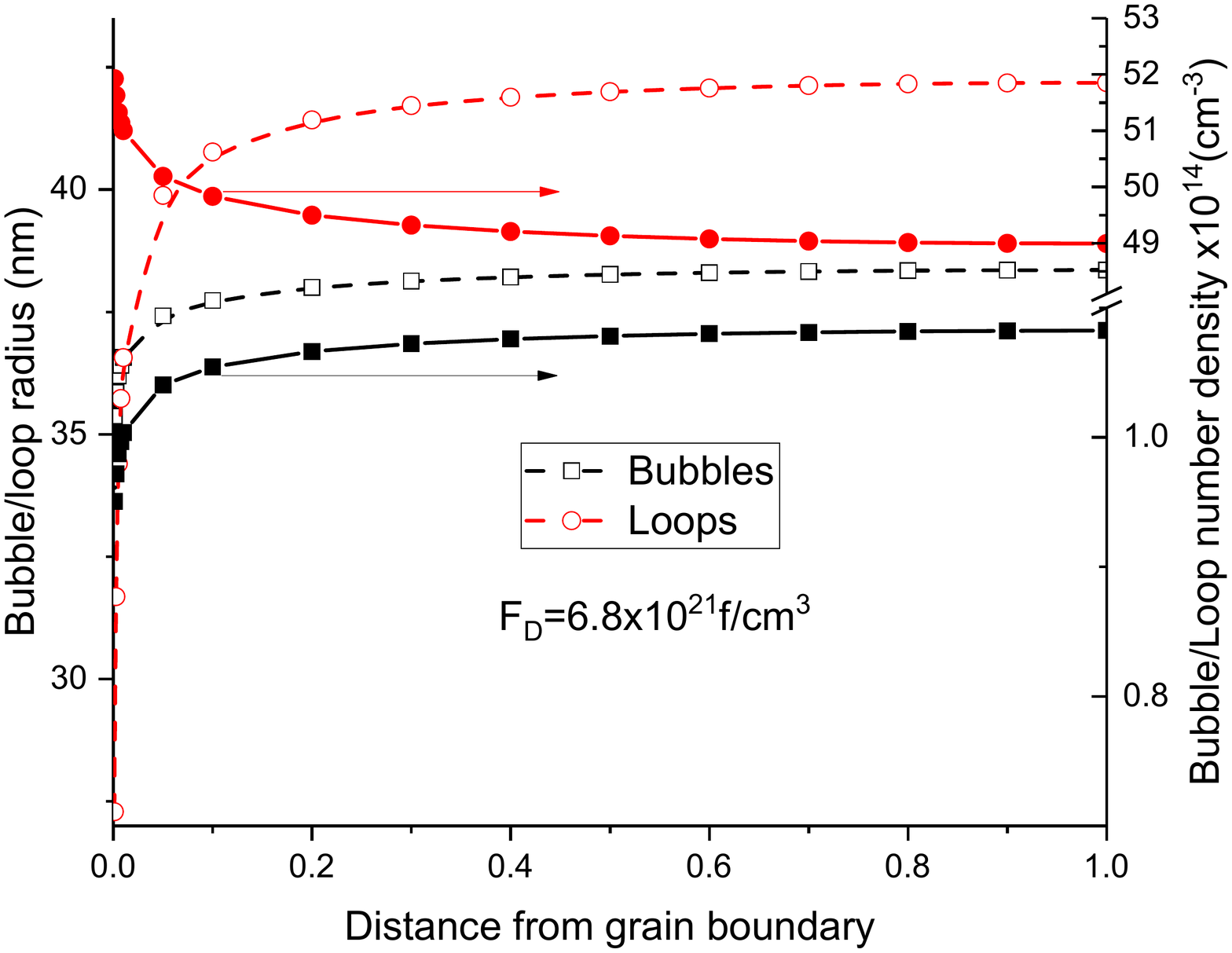}
b)\includegraphics[width=0.4\textwidth]{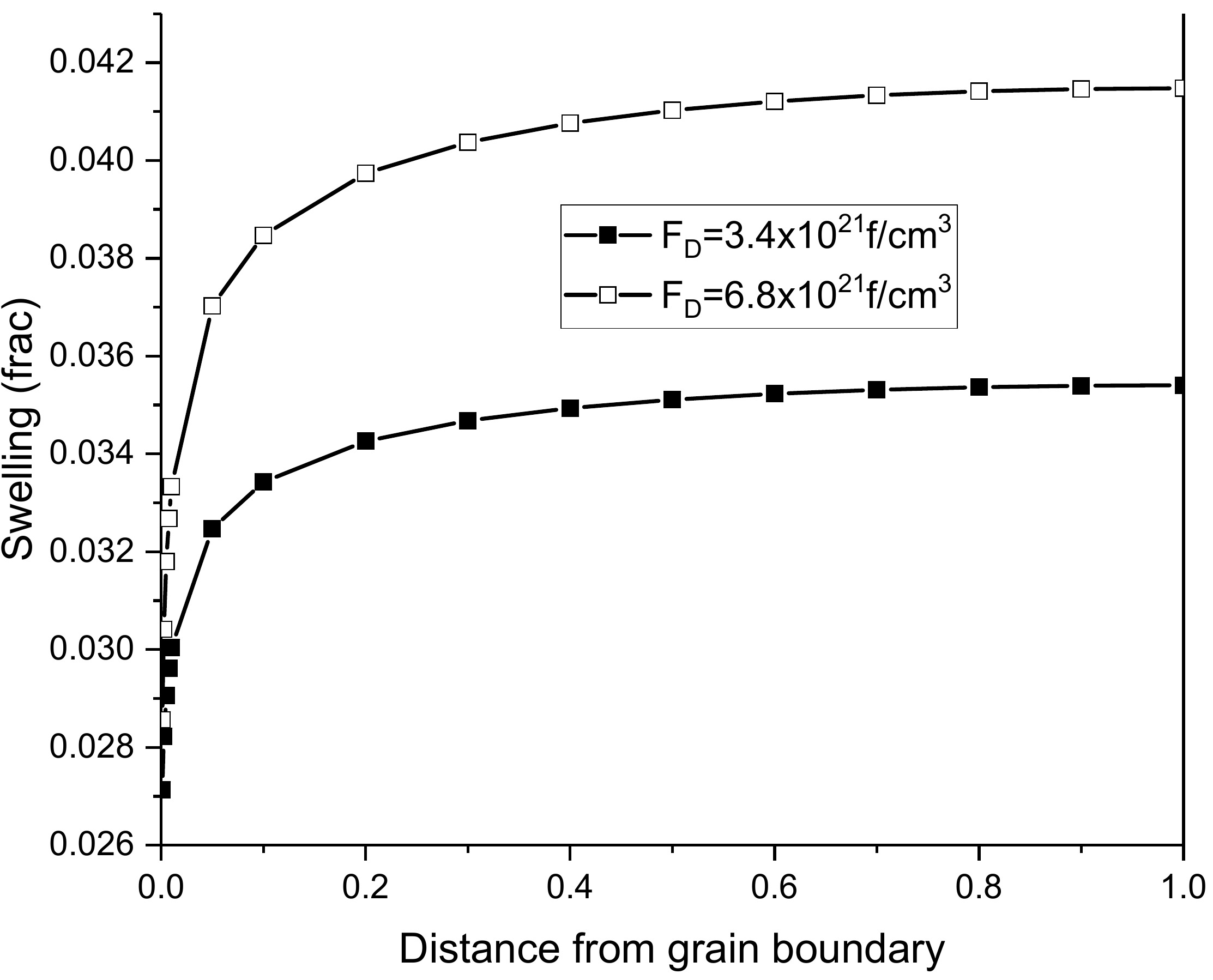}\\
c)\includegraphics[width=0.4\textwidth]{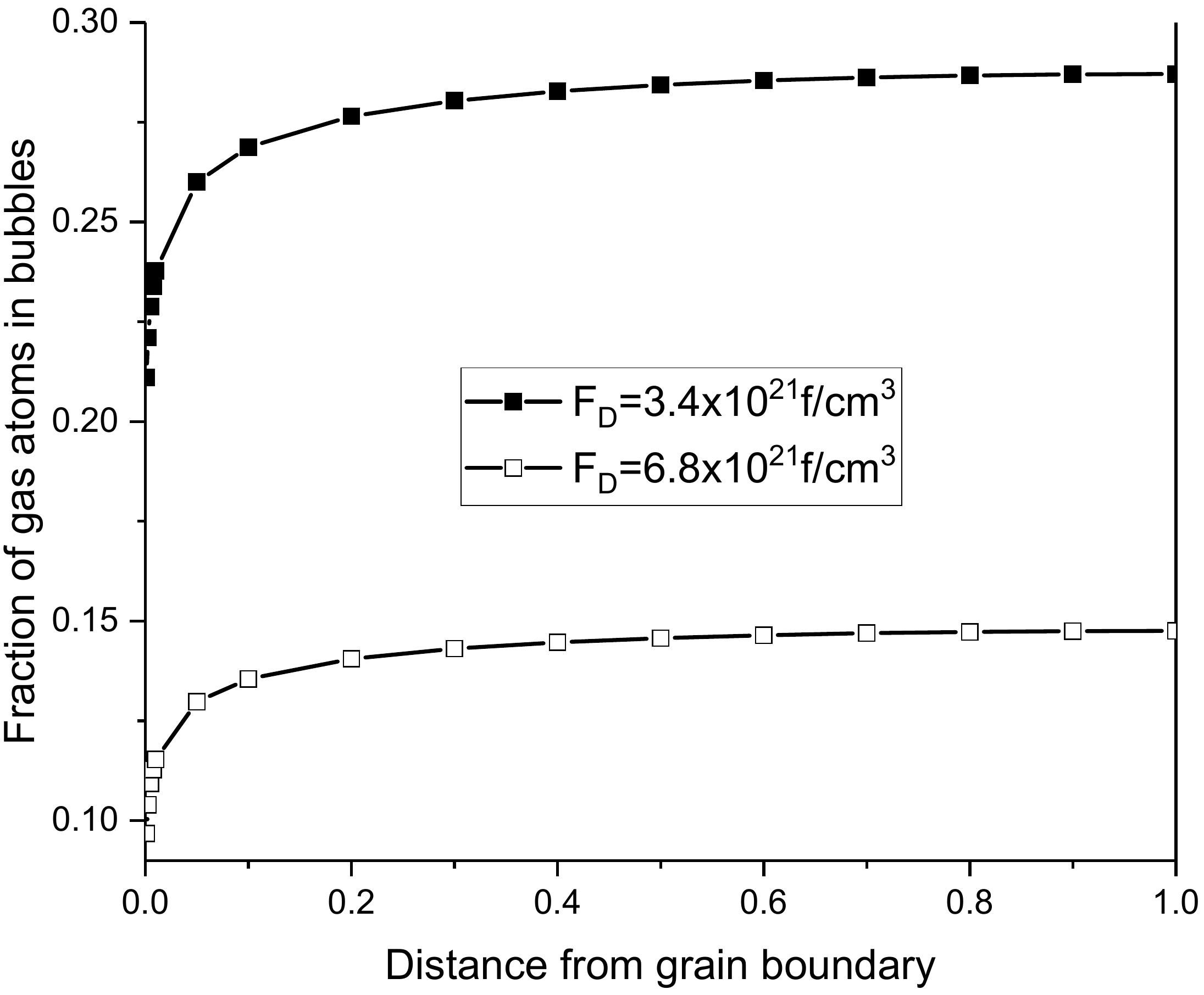}
d)\includegraphics[width=0.4\textwidth]{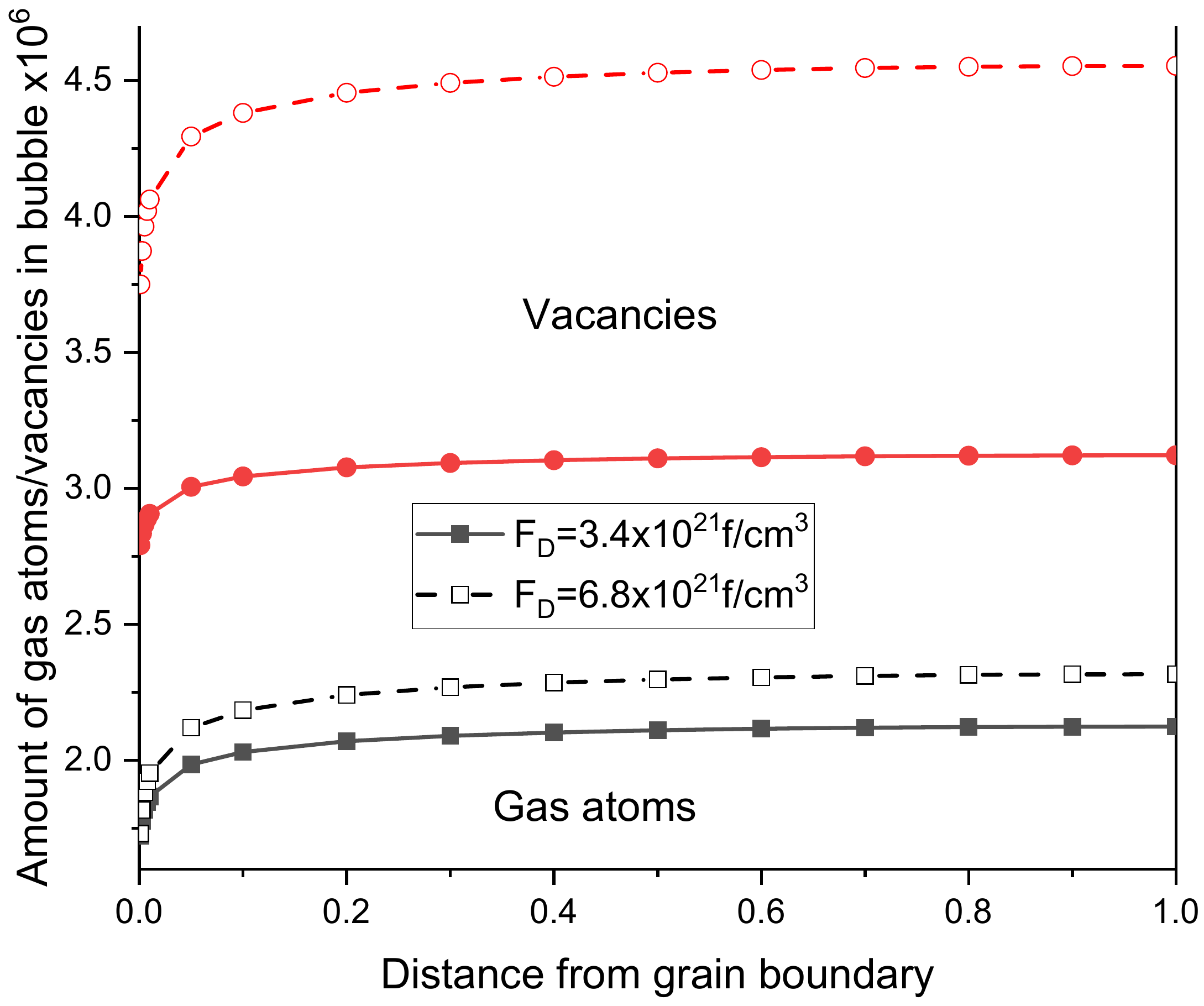}\\
e)\includegraphics[width=0.4\textwidth]{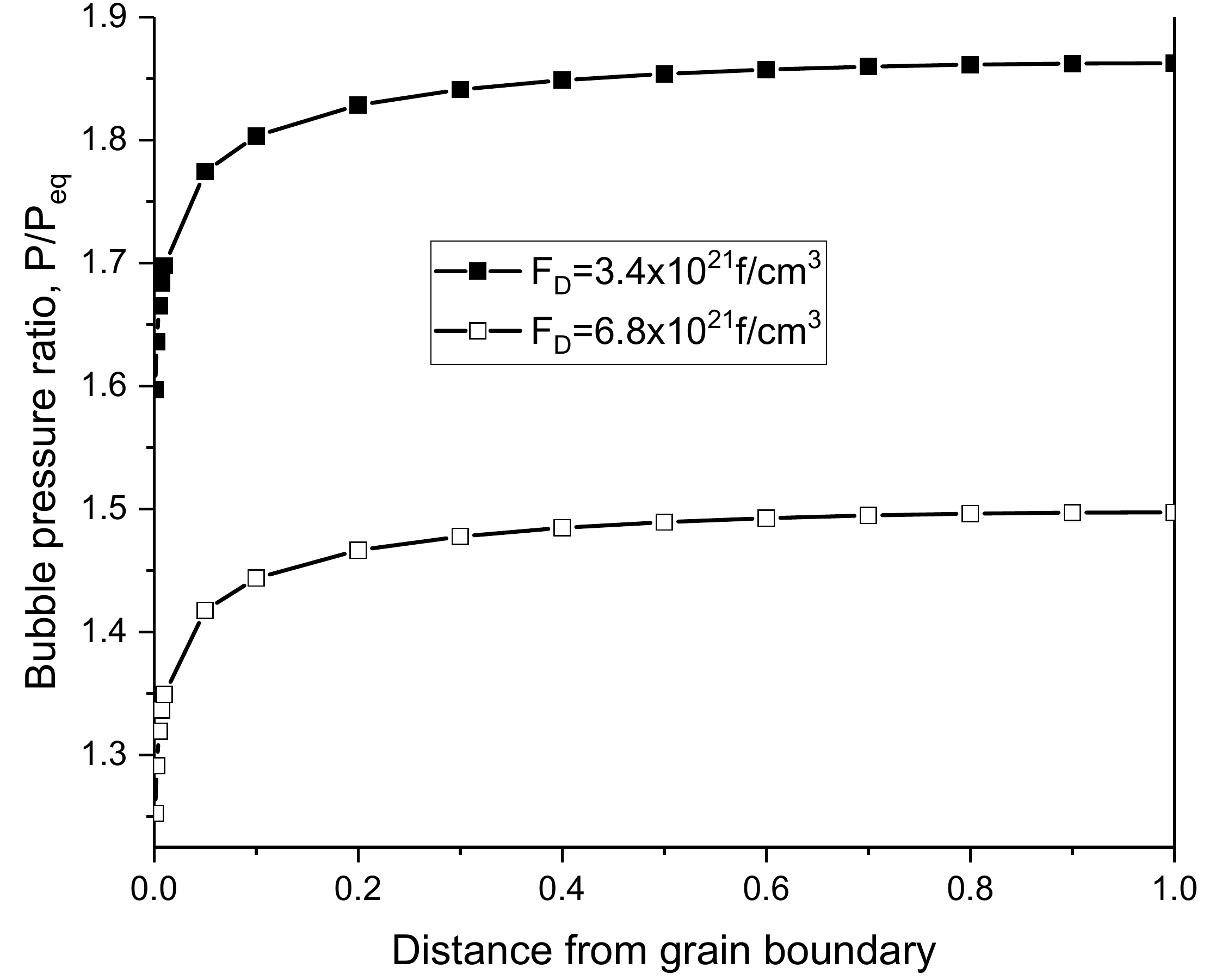}
f)\includegraphics[width=0.4\textwidth]{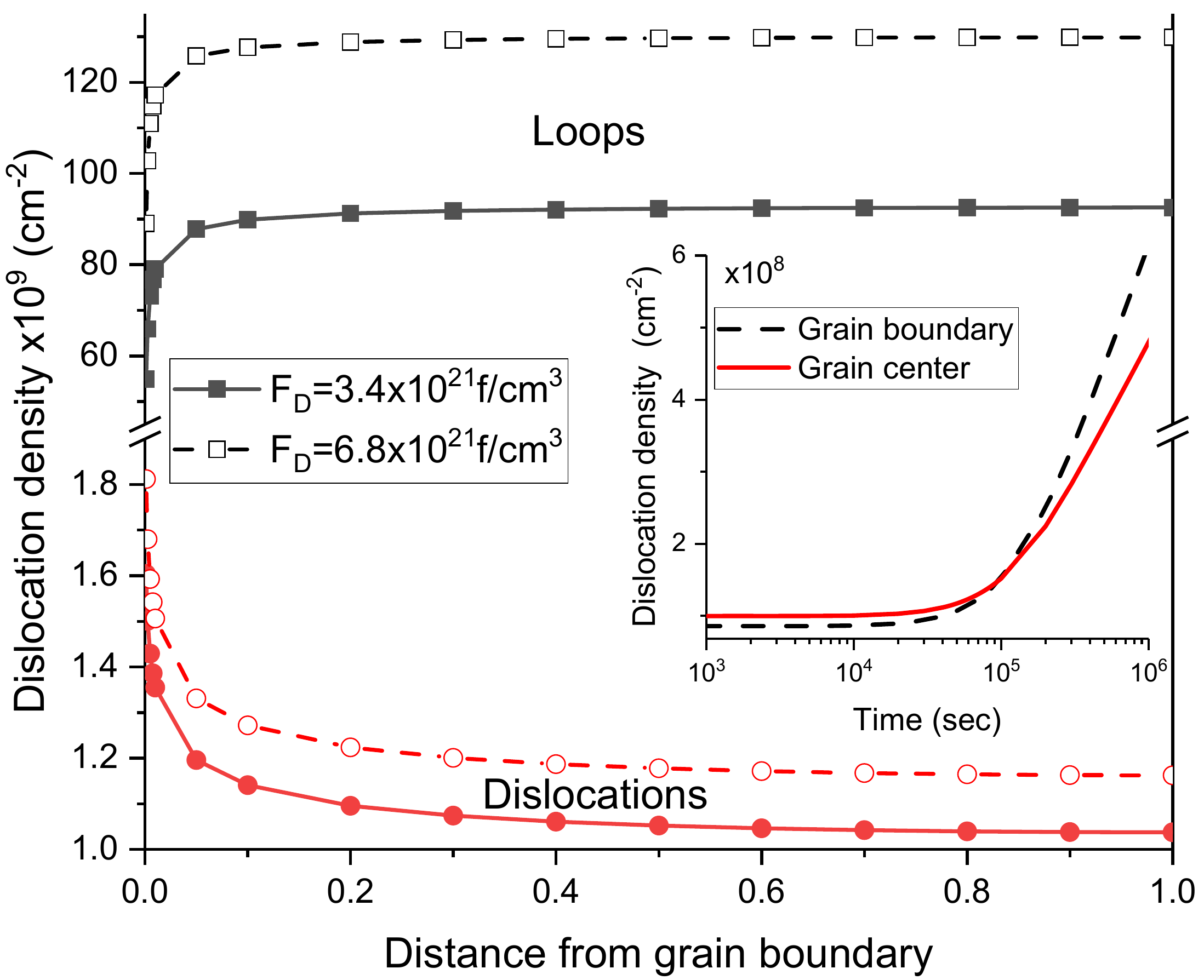}
\caption{Distance dependencies at $T=600K$ and different fission density\label{Ldep1}}
\end{figure}

Let us consider local distribution of bubbles and loops inside grains, see Fig.\ref{Ldep1}. Data  in Fig.\ref{Ldep1}a illustrate that sizes of both bubbles and loops are distributed in nonuniform manner  nearby grain boundary area. The size of bubbles and their number density are higher in domains located in the grain center comparing to their values nearby grain boundaries. From the other hand one gets that small dislocation loops are located with elevated number density nearby grain boundaries, whereas large one will be outside grain boundaries with lower number density.  Such local distribution of both bubbles and loops is caused by these boundaries as extremely sinks with high intensity. From the other hand such a distribution of bubbles inside grains relates to a conclusion that grain will swell inhomogeneously, the swelling is bigger in the grain center where large bubbles grow comparing to domains located nearby grain boundary (see Fig.\ref{Ldep1}b).  By computing fraction of gas atoms in bubbles and  amount of gas atoms and vacancies inside bubbles one finds that these quantities increase in bubbles located closer to the grain center than at the grain boundary (see Fig.\ref{Ldep1}c,d). It is interesting to note that with fission density growth the amount of vacancies in bubbles increases faster than  the number of gas atoms, here the fraction of gas in bubbles decreases with fission density.  
By considering pressure ratio $p/p_{eq}$ in bubbles  at different location (see Fig.\ref{Ldep1}e) one observes that this ratio increases in bubbles located out of grain boundaries, whereas in domains at grain boundaries this ratio goes to unit, with the fission density increase this ratio decreases due to formation of large bubbles enriched by vacancies. Therefore, one concludes that bubbles localized closer to grain boundary are able to emit vacancies, whilst those located nearby grain center will accumulate vacancies.     
By comparing  Fig.\ref{Ldep1}a and Fig.\ref{Ldep1}b one finds that dislocation loop density is lower at grain boundary  comparing to the grain center, despite the number density is higher at grain boundary. In our modeling we admit initially that dislocation density is lower nearby grain boundaries than outside. From obtained data one finds that with the fission density increase the line dislocation density  nearby grain boundaries increases faster than one has for dislocations located far from grain boundaries (see insertion in Fig.\ref{Ldep1}f). As a result, we get a decreasing dependence of line dislocation density when the distance from grain boundary increases at elevated fission density (see circle markers in Fig.\ref{Ldep1}f).

\begin{figure}[!t]
\centering
\includegraphics[width=0.45\textwidth]{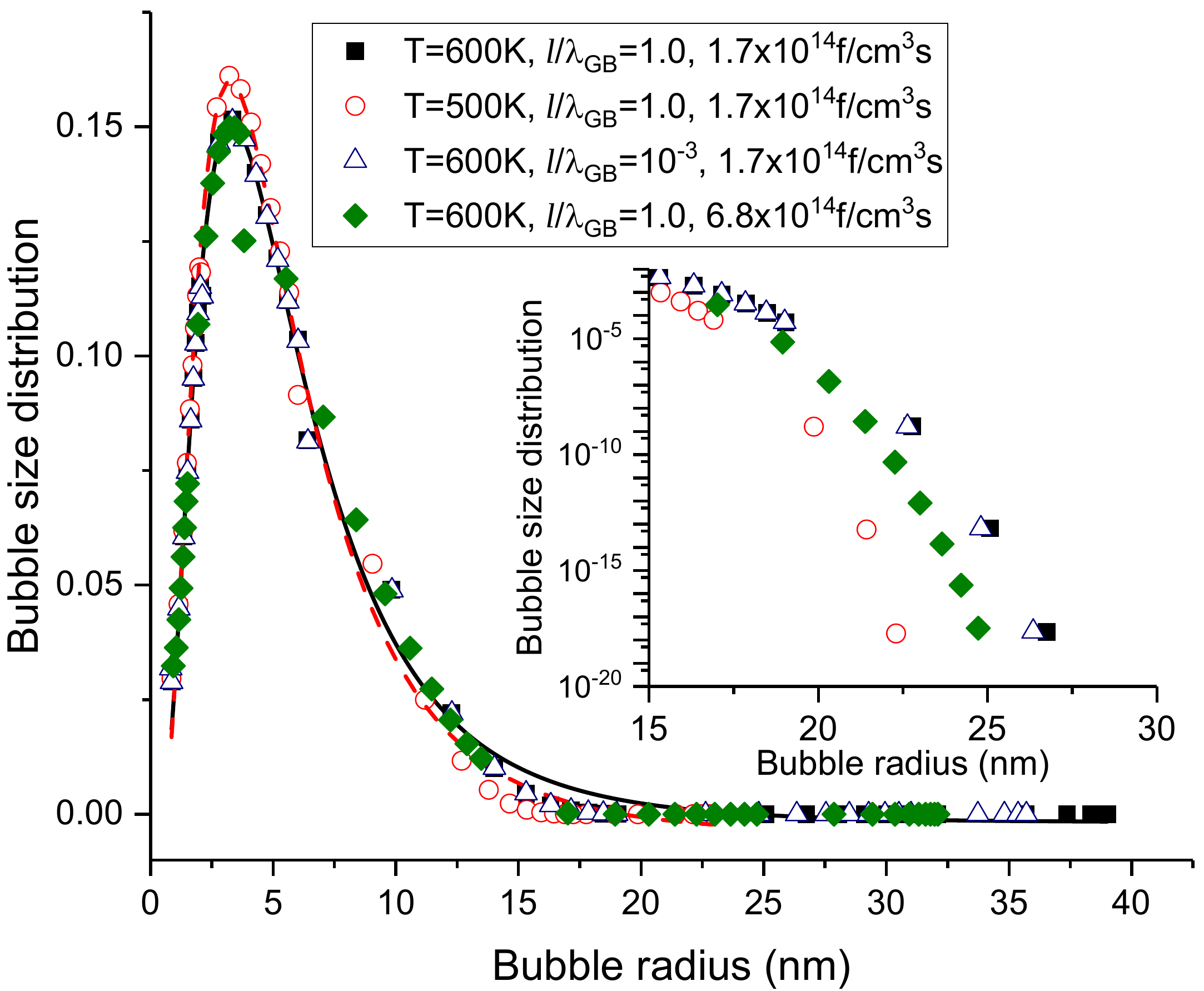}
\caption{Distribution functions over bubble size at different temperature, fission rate, and a distance from grain boundary\label{BSD}}
\end{figure}

Next we pay attention to studying bubble size distribution functions at different temperature, fission rate and  distance from grain boundaries  (see Fig.\ref{BSD}). All obtained data  were normalized and  fitted by Log-Normal distributions and all distributions have the common properties, the difference only in small details. By considering effect of temperature (cf. square and circle markers) one finds that peak of  distribution function remain at the same location, (around 4nm) whereas distribution tail emerges at elevated temperature. The tail is centered at the point approximately equal to five of most probable bubble size ($\approx 20$nm) as was shown in Ref.\cite{Rest2004}. Therefore, at low temperatures there is no so big dispersion in the bubble size. With an increase the fission rate one gets an elevated probability of growing large bubbles (cf. square and diamond markers). Finally, distributions measured at different location from the grain boundary (cf. square and triangle markers) has a difference only in the length of tails, the most of bubbles nearby grain boundaries  are characterized by small size. This effect is caused by affect of grain boundaries as sinks resulting in inhomogeneous distribution of bubbles in the  grain boundary vicinity.     

\section{Comparison with experimental and simulation  data}
The most of irradiation data for $U_3Si_2$ are obtained at temperatures lower than $523K$ (see, for example\cite{Exp2009}), whereas  simulation data relate to temperatures  higher than $550K$ (see Ref.\cite{Barani2019}). In our study we use the relevant temperature interval with the fission rate mostly reported in literature, whilst the discussed fission density relates to 33\% burnup ($U-235$) as was shown in Ref.\cite{Exp2010}. Obtained results in our study are in agreement with the trends of experimental and numerical data known from literature.  

In Ref.\cite{YaoJNM18_498} it was shown that at 84MeV $Xe^+$ irradiation at $573K$ the bubble size was no more than 5nm according to obtained mode size  distributions  at doses from 24dpa up to 80dpa with identified coalescence. In our case a peak of  the size distribution function locates at around 4-5nm at the same temperature range corresponding to experimentally observed results. The difference is observed for bubble number density due to different irradiation sources are exploited.

Obtained data for bubble number density relates well to experimental studies reported in Ref.\cite{Conf2010}, where it was indicated that bubble size in  fuel plate ($U_3Si_2/Al$, $\sim75\%$ enrichment)  irradiated in ATR  at INL varies from 10nm up to 1000nm with the density around $10^{14}cm^{-3}$. At the same time it was shown that at the fuel matrix interaction area the bubble size is small comparing to other areas that relates well to our results of local bubble size distributions. A local gas bubble distribution was shown in Ref.\cite{Exp2009}

In Ref.\cite{Rest2004} it was shown that without geometrical contact bubbles are of nanometer size range with an increased their density at the irradiation. At longer irradiation the density increases in at much-low rate at low temperatures ($T/T_m<0.5$). In our study we observed the same scenario at chosen temperatures. The noted in Ref.\cite{Rest2004} exponential decaying of bubble number density was observed in our modeling and revealed analytically. The fractional swelling  \emph{versus} fission density and its extrapolated values are relate well to theoretical results  and experimental data discussed in Ref.\cite{Rest2004}.   

Temperature dependencies for swelling agrees well to data from GRASS-SST code simulations discussed in Ref.\cite{MKGAYMHY2017} at the studied  temperature interval, where the swelling does not exceed value of 3\%. A good agreement is observed for temperature dependencies of both bubble number density ($(2-9)\times 10^{13}cm^{-3}$, \cite{Finlay04}) and bubble size, where the obtained mean bubble radius  grows from 10nm up to 30nm \cite{Finlay04,Barani2019} with the temperature increase from 550K up to 700K at the fission rate $1.7\times 10^{14}(cm^{-3}s^{-1})$.
Obtained  swelling data agrees with results of Ref.\cite{Conf2019}, the difference in values is related to difference of both  temperature and fission rate. At the same time the trend of swelling at pointed fission density is about 5\%.

Comparison with simulation data discussed in Ref.\cite{Barani2019} illustrates quantitative and qualitative  agreement with data obtained for bubble density and bubble size approximated by a difference in temperature range where simulations were provided. 
%Moreover, despite we use the rate theory approach the formalism discussed in Ref.\cite{Barani2019} exploits simplified dynamical problem but consider different aspects not included in our study.  

\section{Conclusions}

To gain a detailed understanding a negative consequence of fission gas release in light water reactors a modified rate theory model has been used to simulate a growth of fission gas bubbles accompanied with evolving defect microstructure of $U_3Si_2$ fuel. The model is based on  approaches for gas bubble growth  developed  in Refs.\cite{Rest2000,Rest2004}, the set of parameters used in modeling is well justified in previous theoretical and experimental studies reported as standard set in simulation codes such as  GRASS-SST and BISON from IDNL. Our model includes local influence of grain  boundary onto processes of both bubble and microstructure evolution and admits  local distribution of bubbles and dislocation loops/dislocations inside grains.
The developed model  
allows one to monitor statistical and physical  properties of growing bubbles (fraction of gas in bubbles, thermal conductivity, etc.) and estimate local swelling  during fission at different temperature and fission rate with information about defect structure evolution locally inside grains.
To describe universal bubble size distribution a general approach by taking into account gas generation and release is exploited in stationary and auto-model regimes, where universal dynamics of bubble size and its number density at ``knee'' is shown.     

It was shown that growth of fission gas bubbles is mediated by vacancies playing a major role in this process. The amount of vacancies in bubbles increases with temperature and with fission density. Dynamics of bubble size growth is controlled by competition of both vacancies and gas atoms.  We have shown that nucleation stage of bubble size dynamics is changed by coalescence regime. The mean size of bubbles is comparable with the mean size of interstitial dislocation loops and obtained data relate well to predicted values from simulations and experimental observations reported in literature. It was shown that bubble number density manifests two peaks related to small bubbles of the size of several nanometers and large bubbles of several tens nanometers at elevated fission densities. 

It was shown that with the temperature increase from 500K up to 650K with the fission rate from $1.7\times 10^{14}f/cm^3s$ up to $6.8\times 10^{14}f/cm^3s$ the mean bubble size increases with the main contribution coming from vacancies up to 40nm and increasing fraction of fission gas in bubbles. At the same time it was found that an increase in the loop size with temperature up to 45-50nm is associated with decreasing loop number density and loop density. 
We have shown that uniform distribution of both bubbles and loops inside grains is broken in the grain boundary vicinity due to local line dislocations distribution inside grains. 
In the vicinity of grain boundary the size of bubbles is 20-25\% less than  their size in the grain center domains. The loop size reduces by 25-30\% from grain center to boundaries. Statistical distribution of bubbles at different temperature, fission rate, and their locality manifests universal character of Log-Normal distribution  with weak difference in details related to maximal bubble size,  distribution tails, peak position and its height. 
    
%\newpage 

%\addcontentsline{section}{References}
\addcontentsline{toc}{section}{\protect

\end{document}